\documentclass{aa}
\usepackage[varg]{txfonts} 
\usepackage{courier} 
\usepackage{bm} 
\usepackage{upgreek} 
\usepackage{amsmath,amsfonts,amssymb} 
\usepackage{nicefrac} 
\usepackage{mathtools} 
\usepackage[pdfauthor={Alexandros Ziampras, Sareh Ataiee, Wilhelm Kley, Cornelis P. Dullemond, Cl{\'e}ment Baruteau},
pdftitle={The impact of planet wakes on the location and shape of the water iceline in a protoplanetary disk},
pdfsubject={A parameter study using numerical simulations on the impact of planet-driven heating in protoplanetary disks},
pdfkeywords={planet formation, accretion disks, planet--disk interaction, spiral arms, spiral heating, numerical hydrodynamics}]{hyperref} 
\usepackage{subfig} 
\usepackage{float} 
\usepackage{soul} 
\hypersetup{colorlinks=true,citecolor=blue}

\graphicspath{{pics/}} 

    \usepackage{etoolbox}
    \makeatletter
    
    \patchcmd{\NAT@citex}
      {\@citea\NAT@hyper@{%
         \NAT@nmfmt{\NAT@nm}%
         \hyper@natlinkbreak{\NAT@aysep\NAT@spacechar}{\@citeb\@extra@b@citeb}%
         \NAT@date}}
      {\@citea\NAT@nmfmt{\NAT@nm}%
       \NAT@aysep\NAT@spacechar\NAT@hyper@{\NAT@date}}{}{}
    
    \patchcmd{\NAT@citex}
      {\@citea\NAT@hyper@{%
         \NAT@nmfmt{\NAT@nm}%
         \hyper@natlinkbreak{\NAT@spacechar\NAT@@open\if*#1*\else#1\NAT@spacechar\fi}%
           {\@citeb\@extra@b@citeb}%
         \NAT@date}}
      {\@citea\NAT@nmfmt{\NAT@nm}%
       \NAT@spacechar\NAT@@open\if*#1*\else#1\NAT@spacechar\fi\NAT@hyper@{\NAT@date}}
      {}{}
    \makeatother

\newcommand{\tensor}[1]{\overline{\textbf{#1}}}
\newcommand{\tensorGR}[1]{\overline{\bm{{#1}}}}
\newcommand{\DP}[2]{\frac{\partial{#1}}{\partial{#2}}}
\newcommand{\D}[2]{\frac{\mathrm{d}{#1}}{\mathrm{d}{#2}}}

\title{Location and shape of the water iceline in protoplanetary disks}
\title{Planets in radiative protoplanetary disks}
\title{How do planet wakes impact the location and shape of the water iceline in a protoplanetary disk?}
\title{The impact of planet wakes on the location and shape of the water iceline in a protoplanetary disk}

\author{
	Alexandros Ziampras\inst{\ref{inst1}}
	\and Sareh Ataiee\inst{\ref{inst1}}
	\and Wilhelm Kley\inst{\ref{inst1}}
	\and Cornelis P. Dullemond\inst{\ref{inst2}}
	\and Cl{\'e}ment Baruteau\inst{\ref{inst3}}
}

\institute{
	Institut f{\"u}r Astronomie und Astrophysik, Universit{\"a}t T{\"u}bingen, Auf der Morgenstelle 10, 72076 T{\"u}bingen, Germany\label{inst1}
	\and Institute for Theoretical Astrophysics, Zentrum f{\"u}r Astronomie, Heidelberg University, Albert-Ueberle-Str. 2, 69120 Heidelberg, Germany\label{inst2}
	\and Research Institute in Astrophysics and Planetology, University of Toulouse, 14 Avenue Edouard Belin, 31400 Toulouse, France\label{inst3}
}

\date{\today}

\abstract
	{
		Planets in accretion disks can excite spiral shocks, and---if massive enough---open gaps in their vicinity. Both of these effects can influence the overall disk thermal structure.
	}
	{
		We model planets of different masses and semimajor axes in disks of various viscosities and accretion rates to examine their impact on disk thermodynamics and highlight the mutable, non-axisymmetric nature of icelines in systems with massive planets.
	}
	{
		We conduct a parameter study using numerical hydrodynamics simulations where we treat viscous heating, thermal cooling and stellar irradiation as additional source terms in the energy equation, with some runs including radiative diffusion. Our parameter space consists of a grid containing different combinations of planet and disk parameters.
	}
	{
		Both gap opening and shock heating can displace the iceline, with the effects being amplified for massive planets in optically thick disks. The gap region can split an initially hot ($T>170$\,K) disk into a hot inner disk and a hot ring just outside of the planet's location, while shock heating can reshape the originally axisymmetric iceline into water-poor islands along spirals. We also find that radiative diffusion does not alter the picture significantly in this context.
	}
	{
		Shock heating and gap opening by a planet can effectively heat up optically thick disks and in general move and/or reshape the water iceline. This can affect the gap structure and migration torques. It can also produce azimuthal features that follow the trajectory of spiral arms, creating hot zones, ``islands'' of vapor and ice around spirals which could affect the accretion or growth of icy aggregates.
}


\bibpunct{(}{)}{;}{a}{}{,}

\begin{document}
	\maketitle
	\section{Introduction}
	\label{section:intro}
	
	Protostellar disks are the birth sites of all sorts of planets. Several observations such as the discovery of PDS 70b \citep{keppler-etal-2018} and the recent DSHARP survey \citep{andrews-etal-2018} have spatially resolved such disks, providing valuable constraints on their composition, structure, and possible planets they might harbor. Dust continuum observations reveal annular structures and non-axisymmetric features such as spirals, crescents, or blobs, all of which are consistent with the planet formation scenario \citep{zhang-etal-2018}. According to this scenario, a sufficiently massive planet can trap dust particles by forming pressure maxima \citep[e.g.,][]{ataiee-etal-2018} at radii close to its semimajor axis as it launches density waves in the form of spiral arms that permeate the disk \citep{ogilvie-lubow-2002}. These pressure traps can allow dust particles to concentrate enough for their emission to be observable, and also provide an environment for them to collide and grow.
	
	Dust growth is expected to be further facilitated around opacity transition regions \citep{drazkowska-alibert-2017,zhang-etal-2015}. Common dust opacity models are in principle density- and temperature-dependent---with boundaries defined at conditions where aggregates of certain composition change phase---such that crossing between two opacity regimes can change the absorption/emission properties of the disk. For example, the water content of ice-coated particles sublimates at the so-called \emph{water iceline} around $T_\mathrm{ice}\approx170$\,K \citep{lin-papaloizou-1985}, with small variations depending on model assumptions. This temperature marks the first opacity transition threshold that particles will cross as they drift inwards according to several opacity models \citep[e.g.,][]{bell-lin-1994,semenov-etal-2003}. Since water can only be found on particles outside of this iceline, its location can provide insight and constraints on the origin of water content of planetesimals and young planets in an evolving protostellar disk, depending on the disk's temperature profile \citep{bitsch-etal-2019}.
	
	The disk's thermal structure depends on the balance between heating and cooling terms. 
	\citet{kley-crida-2008} showed that accounting for radiation transport instead of treating the disk as locally isothermal can have significant effects on the migration of super-Earths by slowing down or even reversing the migration rate. Additionally, \citet{rafikov-2016} showed that shock heating due to planet-induced spirals can be a significant heat source in the inner few au of the disk. Evidently, the optically thick region near the star can reach high densities and temperatures, which could lead to an important contribution by shock heating to the energy content of the disk.
	
	In this study, we investigate the conditions under which planet shock heating can significantly raise temperatures, and the degree that spirals can affect the location and shape of the water iceline. Based on our findings, we speculate about possible implications on dust and planetesimal growth around the iceline.

	In Sect.~\ref{section:shock-estimate} we calculate an estimate for the amount of heat a planet can pump into the disk through shock heating. Our physical framework as well as numerical setup is described in Sect.~\ref{section:model}. In Sects.~\ref{section:profiles}~and~\ref{section:icelines} we present our results regarding the disk structure and iceline shape, respectively. In Sect.~\ref{section:discussion} we comment on our findings and discuss their potential implications, while Sect.~\ref{section:conclusions} contains a summary of our work along with our conclusions.

	\section{Shock heating}
	\label{section:shock-heating}
	
	Planet-induced spiral arms form as a result of density waves shearing in the Keplerian disk flow as they propagate away from the planet \citep{kley-nelson-2012}. They are overdensities with respect to the disk ``background" (azimuthally-averaged) profile that can steepen into shocks as the travel through the disk \citep{goodman-rafikov-2001}. In an adiabatic framework, we can expect a pressure jump at the location of the shock, which can lie close to the planet \citep{zhu-etal-2015}. This pressure jump can generate heat near the planet, potentially affecting the temperature profile near the corotating region. The question then is, how important this shock heating can be when compared to other heat sources in the disk (e.g., viscosity and stellar irradiation).
		
	In order to get a clue about the prominence of spiral shocks as a heat-generating mechanism, it is worthwhile to first estimate their contribution theoretically and compare to other sources of heat in the disk. We follow a line of thought similar to \citet{rafikov-2016} and calculate the heating by an adiabatic spiral shock for an assumed density jump at the shock. 
	
	\label{section:shock-estimate}
	
	The heating by a spiral shock can be considered as a three-phase process: (1)~heating by the shocks; (2)~decompression; and (3)~settling to the pre-shock density. In this subsection, we will refer to the pre-shock quantities (before phase 1) by the subscript $1$, to the decompression phase by subscript 2, and to the post-shock state by subscript $3$. The following calculation is performed in the shock's comoving frame. The shock heating rate can be estimated by calculating the specific internal energy difference between phase~1 and phase~3 for each passage of the shock and then dividing it by the time between two passages. Knowing the pressure $p$ and surface density $\Sigma$ in each of the three phases, we can calculate the specific internal energy via $e = p/(\Sigma (\gamma -1))$. The classical jump condition and equation of state can give us the values of all needed quantities. In our calculations, we use the surface density $\Sigma=\int_{-\infty}^{+\infty}\rho \mathrm{d}z$ instead of the density $\rho$. Note that the jump conditions are also valid if $\rho$ is replaced by $\Sigma$ because, during the shock, the disk does not have enough time to expand vertically and change the local density. This allows us to use two-dimensional (2D) hydrodynamic simulations to test the predictions of this analytical model.
	
\begin{figure}
		\centerline{\includegraphics[width=0.5\textwidth]{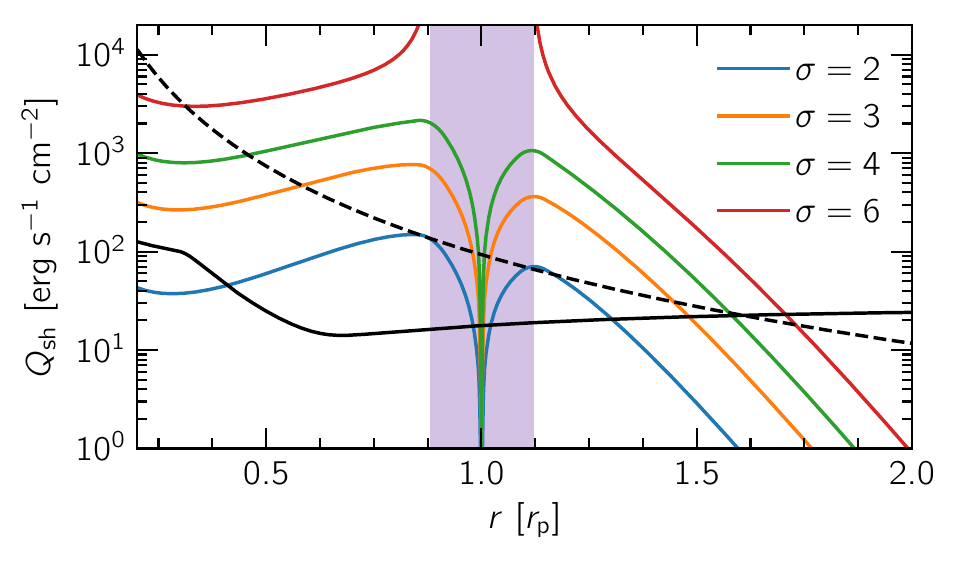}}
		\caption{The shock heating rate by the planet's spirals estimated by the method in Sec.~\ref{section:shock-estimate} and damped using Eq.~\eqref{eq:shockstrength-damp} for different shock strengths ($\sigma=\Sigma_2/\Sigma_1$, $\gamma=\nicefrac{7}{5}$) compared to the viscous and irradiative heating rates (dashed and solid black lines; see Sect.~\ref{section:physics}). The lilac band indicates the corotating region, in which the estimates are not valid due to potential gap opening. The model used for this plot assumes $M_\mathrm{p}=100$\,$\mathrm{M_\oplus}$, $\dot{M}=10^{-8}$\,$\mathrm{M_\odot/yr}$, $\alpha=10^{-3}$, $r_\mathrm{p}=4$\,au (see Sects.~\ref{section:physics},~\ref{section:numerics}).}
		\label{fig:heatingrates}
	\end{figure}

	{\em Phase~1 $\rightarrow$ phase~2}: When the shock hits the pre-shock gas, the density and pressure at the second phase can be given by the Rankine--Hugoniot jump condition that reads
	\begin{eqnarray}
	\frac{\Sigma_2}{\Sigma_1} &=& \frac{(\gamma+1) \mathcal{M}_{1}^2}{(\gamma-1)\mathcal{M}_{1}^2+2},
	\label{eq:jumpcondition1}
	\\
	\frac{p_2}{p_1} &=& \frac{2 \gamma \mathcal{M}_{1}^2-(\gamma-1)}{\gamma+1}
	\label{eq:jumpcondition2}
	\end{eqnarray}
	\noindent where $\gamma$ is the adiabatic index and $\mathcal{M}$ denotes the Mach number.
	\\
	{\em Phase~2 $\rightarrow$ phase~3}: The decompression phase can be either adiabatic or isothermal depending on disk cooling. If the shocks are in the optically thin part of the disk, the inserted energy can be easily radiated away, the post-shock temperature returns to its pre-shock value rapidly (except in a very narrow region of the shock itself), and the decompression would be isothermal. 
	Conversely, in the optically thick part of the disk where the energy cannot quickly escape, the decompression is adiabatic. Because we are interested in the cases where shocks heat the disk up and change its temperature structure, we choose the adiabatic decompression. Therefore, the pressure and density before and after decompression can be given by an adiabatic equation of state as
	\begin{equation}
		\label{eq:decompression1}
		\frac{p_3}{p_2}= \left( \frac{\Sigma_3}{\Sigma_2} \right)^{\gamma}.
	\end{equation}
	\noindent Because the gas density will return to its pre-shock value, we can replace 
	$\Sigma_3$ with $\Sigma_1$ so that:
	\begin{equation}
		\label{eq:decompression2}
		p_3 = p_2 \left(\frac{\Sigma_1}{\Sigma_2}\right)^{\gamma}.
	\end{equation}
	\\
	Let's assume that the time between two passages of a shock through a specific location of radius $r$ in the disk is $t_{\mathrm{pass}}=2\pi/|\Omega_K(r)-\Omega_K(r_\mathrm{p})|$, where $\Omega_\mathrm{K}=\sqrt{\mathrm{G}(M_\ast+M_\mathrm{p})/r^3}$ is the keplerian frequency, $\mathrm{G}$ the gravitational constant, $M_\ast$ and $M_\mathrm{p}$ the masses of the star and planet respectively, and $r_\mathrm{p}$ the planet's semimajor axis. The amount of heat per unit time (averaged over many passages) is then:
	\begin{eqnarray}
	\begin{split}
		Q_{\rm sh} &=\frac{\Delta (\Sigma e)}{\Delta t}=\frac{e_3-e_1}{t_{\rm pass}} \Sigma_{1} = \frac{1}{t_{\rm pass}(\gamma-1)}\left[p_2 \left(\frac{\Sigma_1}{\Sigma_2}\right)^{\gamma} - p_1 \right].
	\end{split}
	\end{eqnarray}
	Expressing $\mathcal{M}_1$ from Eq.~\eqref{eq:jumpcondition1} we obtain:
	\begin{equation}
		\label{eq:Mach}
		\mathcal{M}_{1}^2 = \frac{2 (\Sigma_2/\Sigma_1)}{(\gamma+1)-(\gamma-1)(\Sigma_2/\Sigma_1)}.
	\end{equation}
	Inserting this into Eq.~\eqref{eq:jumpcondition2}, we can remove $p_2$ from the above equation and obtain:
	\begin{equation}
	\label{eq:Qshock}
	Q_{\rm sh}=
	\frac{p_1}{t_\mathrm{pass}(\gamma-1)} \times \left[\left(\frac{1}{\sigma}\right)^{\gamma} \frac{(\gamma+1)\sigma- (\gamma-1)}{(\gamma+1)-(\gamma-1)\sigma} -1\right],
	\end{equation}
	where $\sigma\coloneqq\Sigma_2/\Sigma_1$ is the ``shock strength''. This equation is identical to Eq.~16 of \cite{rafikov-2016} assuming a one-armed spiral. In the literature, there is no straightforward way to find the strength of planetary spirals. Following \citet{rafikov-2016}, we take $\sigma$ as a free parameter and compare the shock heating rate with the viscous and the irradiation heating rates. This comparison is shown in Fig.~\ref{fig:heatingrates}, where we artificially damp $\sigma$ exponentially with distance from the planet's gap opening region to avoid overestimating shock heating far from the planet:
	\begin{equation}
	\label{eq:shockstrength-damp}
	\sigma(r) =
		\begin{cases} 
		\sigma, & |r-r_\mathrm{p}|\leq 2.5\,\mathrm{R_{Hill}}, \\
		1+(\sigma-1)e^{-4(|r-r_\mathrm{p}|-2.5\,\mathrm{R_{Hill}})/r_\mathrm{p}}, & \text{otherwise},
		\end{cases}
	\end{equation}
	where $\mathrm{R_{Hill}}\equiv \sqrt[3]{M_\mathrm{p}/3M_\ast}\,r_\mathrm{p}$.
		
	For $M_1^2\rightarrow\infty$, Eq.~\eqref{eq:jumpcondition1} gives an upper limit to shock strength for adiabatic shocks as $\sigma\rightarrow\frac{\gamma+1}{\gamma-1}=6$ for $\gamma=\nicefrac{7}{5}$. We should note that this upper limit is not strict if additional thermal mechanisms (such as cooling) are included in the models.
	
	This estimate shows that shock heating by a planet can overcome the other two heating sources if the planet is massive enough to produce strong shocks, and the disk opacity is large enough to prevent heat from quickly escaping from the midplane. This extra heating raises the temperature in the disk up to the location where the shocks damp greatly. Because the disk opacity also depends on temperature (see Fig.~\ref{fig:opacity}), spiral heating by a planet in the vicinity of an iceline (either via migration or in-situ formation) might displace the latter. In the following sections we study this problem for a more realistic model with shocks that are not necessarily adiabatic, and examine how much and under which conditions the location and shape of an iceline can change.
	
	\section{Model setup}
	\label{section:model}
	In this section, we present the physical framework that we utilize in our planet--disk modeling. We list relevant equations, the assumptions behind them, and describe our numerical setup as far as our parameter space, initial/boundary conditions and grid structures are concerned.
	
		\subsection{Physics}
		\label{section:physics}
		
		We solve the vertically integrated Navier--Stokes equations for a disk with surface density $\Sigma$, velocity vector $\bm{v}$ and vertically integrated specific internal energy $e$ on a polar coordinate system $\{r,\phi\}$ centered around the star. For a perfect gas the equations read
		\begin{equation}
			\label{eq:navier-stokes}
			\begin{split}
				\D{\Sigma}{t}&=-\Sigma\nabla\cdot\bm{v},\\
				\Sigma\D{\bm{v}}{t}&=-\nabla p-\Sigma\nabla\Phi+\nabla\cdot\tensorGR{\upsigma},\\
				\D{\Sigma e}{t}&=-\gamma\Sigma e\nabla\cdot\bm{v}+Q_\mathrm{visc}+Q_\mathrm{irr}-Q_\mathrm{cool},
			\end{split}
		\end{equation}
		where $\gamma=\nicefrac{7}{5}$ is the adiabatic index, $p=(\gamma-1)\Sigma e$ is the vertically integrated pressure and $\tensorGR{\upsigma}$ denotes the viscous stress tensor.
		
		The adiabatic and isothermal sound speeds $c_\mathrm{s}$, ${c_\mathrm{s}}_\mathrm{iso}$ are related as:
		\begin{equation}
			{c_\mathrm{s}}_\mathrm{iso}=c_\mathrm{s}/\sqrt{\gamma}=H\Omega_\mathrm{K}=h v_\mathrm{K}=\sqrt{\mathrm{R}T/\mu},
		\end{equation}
		where $H$ is the pressure scale height, $h=H/r$ is the aspect ratio and $v_\mathrm{K}=\Omega_\mathrm{K} r = \sqrt{\mathrm{G}M/r}$ is the Keplerian azimuthal velocity. The universal gas constant and mean molecular weight are denoted by $\mathrm{R}$ and $\mu=2.353$, respectively.
		
\begin{figure}[h]
			\centering
			\includegraphics[width=.5\textwidth]{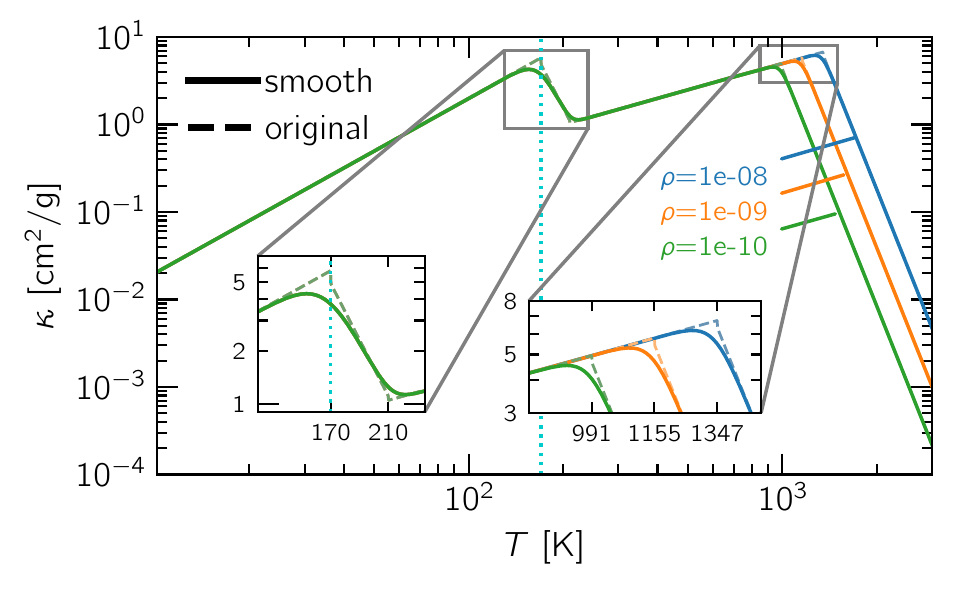}
			\caption{First 4 opacity regimes according to \citet{lin-papaloizou-1985}. The dotted teal line marks the water iceline ($T_\mathrm{ice}=170$\,K). Different branches are patched together by interpolation.}
			\label{fig:opacity}
		\end{figure}

		Source terms $Q_\mathrm{visc}$, $Q_\mathrm{irr}$, and $Q_\mathrm{cool}$ in the energy equation correspond to viscous heating, stellar irradiation, and thermal cooling, respectively:
		\begin{equation}
			\label{eq:source-terms}
			\begin{split}
				Q_\mathrm{visc}&=\frac{1}{2\nu\Sigma}\left(\sigma_{rr}^2+2\sigma_{r\phi}^2+\sigma_{\phi\phi}^2\right)+\frac{2\nu\Sigma}{9}\left(\nabla\cdot\bm{v}\right)^2,\\
				Q_\mathrm{irr}&=2\frac{L_\star}{4\mathrm{\pi} r^2}(1-\epsilon)\left(\D{\log H}{\log r}-1\right)h\frac{1}{\tau_\mathrm{eff}},\\
				Q_\mathrm{cool}&=2\sigma_\mathrm{SB}\frac{T^4}{\tau_\mathrm{eff}},\\
			\end{split}
		\end{equation}
		where $\nu=\alpha c_\mathrm{s} H$ is the kinematic viscosity according to the $\alpha$-viscosity model of \citet{shakura-sunyaev-1973}, $\sigma_\mathrm{SB}$ is the Stefan--Boltzmann constant, and $\tau_\mathrm{eff}$ is an effective optical depth following \citet{hubeny-1990}:
		\begin{equation}
			\tau_\mathrm{eff}=\frac{3\tau}{8}+\frac{\sqrt{3}}{4}+\frac{1}{4\tau},\qquad \tau=\int_{0}^{\infty}\kappa\rho dz\approx c_1\kappa\rho_\mathrm{mid}H,
		\end{equation}
		with the Rosseland mean opacity $\kappa(\rho,T)$ defined according to \citet{lin-papaloizou-1985}, shown in Fig.~\ref{fig:opacity}. 
		The correction factor~$c_1=\nicefrac{1}{2}$ is added following \citet{mueller-kley-2011} to account for the drop in opacity with height. We assume a Gaussian vertical density profile so that $\Sigma=\sqrt{2\mathrm{\pi}}\rho_\mathrm{mid}H$.
		
		As far as irradiation is concerned, we assume a star of solar luminosity $L_\ast=\mathrm{L}_\odot$ and a disk albedo of $\epsilon=\nicefrac{1}{2}$. Following \citet{menou-goodman-2004}, the factor $\D{\log H}{\log r}$ is assumed to be constant and equal to $\nicefrac{9}{7}$ (i.e., disk self-shadowing is not considered).
		
		\subsection{Numerics}
		\label{section:numerics}
		
	    We utilize the numerical MHD code \texttt{PLUTO} \citep{mignone-etal-2007} for our simulations, along with the FARGO algorithm described by \citet{masset-2000} and implemented as a library into \texttt{PLUTO} by \citet{mignone-etal-2012}. To enable radiative diffusion, we implemented a separate module that is briefly described in Appendix~\ref{appendix:raddiff}. Simulations with an embedded planet run on a polar $\{r,\phi\}$ grid, logarithmically-spaced in the radial direction.
		    
		Our parameter space is shown in Table~\ref{table:parameter-space}. It contains the planet mass $M_\mathrm{p}$, the planet's semimajor axis $r_\mathrm{p}$ (fixed, circular orbits), the viscosity parameter $\alpha$ and the initial disk mass accretion rate $\dot{M}$, which is constant throughout the disk in viscous equilibrium such that $\dot{M}=3\pi\nu\Sigma$ \citep{lodato-2008}. By selecting an $\alpha$ value and a constant accretion rate we can then construct well-defined equilibrium states.
		    
		\begin{table}[h]
			\centering
			\caption{Parameter space: $r_\mathrm{p}$ is quoted in au and $\dot{M}$ in $\mathrm{M}_\odot$/yr.}
			\begin{tabular}{r|c}
				\hline
				 parameter & values\\
				\hline
				\hline
				$M_\mathrm{p}$ & $10$\,$\mathrm{M}_\oplus$, $100$\,$\mathrm{M}_\oplus$, $1$\,$\mathrm{M}_\mathrm{J}$, $3$\,$\mathrm{M}_\mathrm{J}$\\
				$r_\mathrm{p}$ & $1$, $4$, $10$\\
				$\dot{M}$ & $10^{-9}$, $10^{-8}$, $10^{-7}$\\
				$\alpha$ & $10^{-4}$, $10^{-3}$, $10^{-2}$\\
				\hline
			\end{tabular}
			\label{table:parameter-space}
		\end{table}
		    
\begin{figure}[h]
		    	\centering
		    	\includegraphics[width=.5\textwidth]{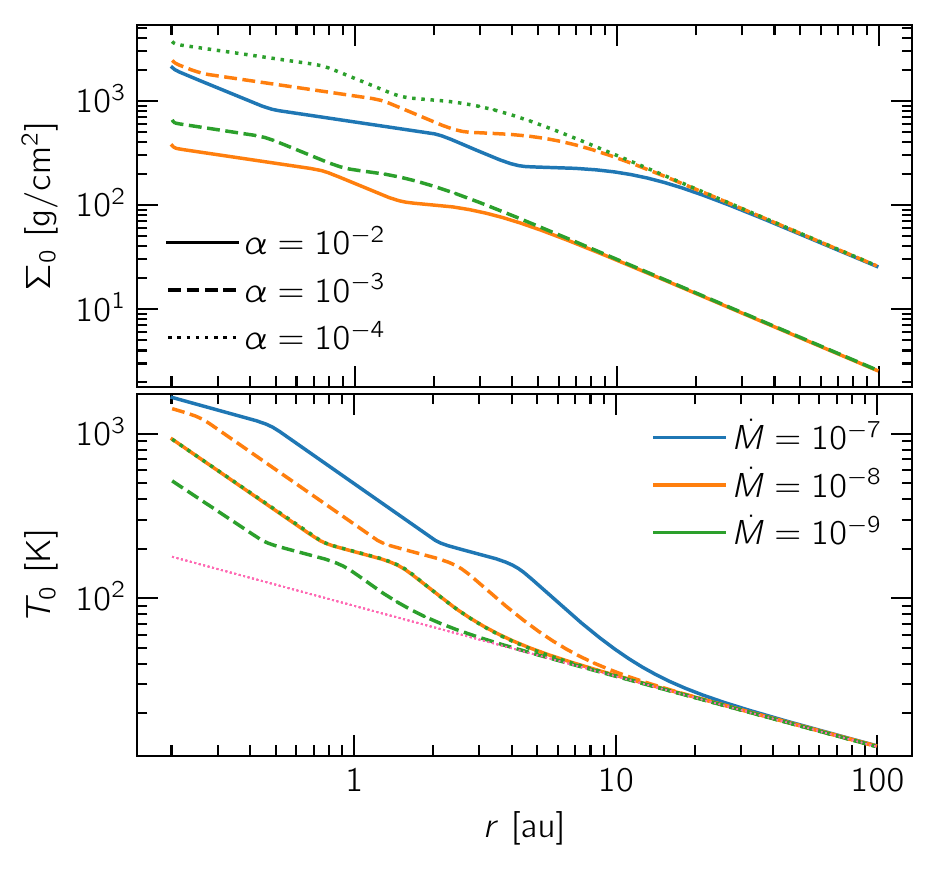}
		    	\caption{Initial profiles for the 5 disk models with different $\dot{M}$ and $\alpha$ that are used throughout the study. The dotted pink line refers to an inviscid disk where $Q_\mathrm{cool}=Q_\mathrm{irr}$ (i.e., the irradiation temperature) and functions as our effective temperature floor. It becomes clear that viscous heating is strongest in the inner disk, while irradiation dominates its outer parts.}
		    	\label{fig:initial-profiles}
		    \end{figure}

		    To generate our initial conditions, we prepare 1D models that satisfy viscous and thermal equilibrium conditions:
		    \begin{equation}
		    	\begin{split}
		    		\dot{M}&=3\pi\nu\Sigma\qquad\qquad\text{\phantom{e}(viscous equilibrium)},\\
		    		Q_\mathrm{cool}&=Q_\mathrm{visc}+Q_\mathrm{irr}\qquad\text{(thermal equilibrium)},
		    	\end{split}
		    \end{equation}
		    and rule out very cold disks or gravitationally unstable ones, for which the Toomre parameter $Q_\mathrm{T}$ \citep{toomre-1964}, defined as:
		    \begin{equation}
		    	Q_\mathrm{T} \equiv \frac{c_\mathrm{s}\Omega_\mathrm{K}}{\mathrm{\pi}\mathrm{G}\Sigma},
		    \end{equation}
		    is less than unity. The initial profiles used are plotted in Fig.~\ref{fig:initial-profiles}.
		    
		    We then embed planets in each configuration and run until the disk roughly reaches viscous and thermal equilibrium or a maximum simulated time of $t_\mathrm{max} = 10^5$ years elapses. To ensure a constant $\dot{M}$ through the boundaries, $\Sigma$, $\bm{v}$ are damped to the initial profiles according to \citet{devalborro-etal-2006} over a timescale of 0.3 boundary orbital periods.
		    
		    The gravitational forces read
		    \begin{equation}
		    \begin{split}
		    	\bm{g}_\mathrm{grav}&=
		    	\bm{g}_\mathrm{\ast}+\bm{g}_\mathrm{p}+\bm{g}_\mathrm{in}=-\nabla\Phi\\
		    	&=
		    	-\frac{\mathrm{G}M_\ast}{r^3}\bm{r}
		    	-\frac{\mathrm{G}M_\mathrm{p}}{(r_\mathrm{e}^{2}+\epsilon^2)^{3/2}}\bm{r}_\mathrm{e}-\frac{\mathrm{G}M_\mathrm{p}}{r_\mathrm{p}^3}\bm{r}_\mathrm{p},\quad \bm{r}_\mathrm{e}=\bm{r}
		    	-\bm{r}_\mathrm{p},
		    \end{split}
		    \end{equation}
		    where $\bm{g}_\mathrm{\ast}$, $\bm{g}_\mathrm{p}$, $\bm{g}_\mathrm{in}$ refer to the gravitational acceleration by the star, the planet, and the indirect term that arises due to the star--planet system orbiting around their mutual barycenter.
		    The planet is on a fixed orbit and we neglect the backreaction of the disk onto star and planet.
		    For the softening length we use $\epsilon=0.6H$ to prevent singularities around the planet's location. The value 0.6 is selected according to \citet{mueller-etal-2012} as it provides very similar results to 3D models. We note that $\epsilon$ is evaluated using the local $H$ at each cell.
		   
		\section{Disk profiles}
		\label{section:profiles}

		Having described our physics and numerical methods, we proceed to execute our simulations.  The grid setup for each model is shown in Table~\ref{table-grid}. A cross-code comparison as well as a resolution test for the verification of our numerical setup is provided in Appendix~\ref{appendix:verification}. We constructed the numerical grid such that the pressure scale height $H$ is resolved by at least 6 grid cells at the planet's location (see Appendix.~\ref{appendix:grid-structure} for more details).
				
		\begin{table}[h]
			\centering
			\begin{tabular}{c|c|c||c||c|c}
				\hline
				$\log\left(\frac{\dot{M}}{\mathrm{M_\odot}/\mathrm{yr}}\right)$ & $\log\alpha$ & $r_\mathrm{p}$ [au] & $M_\mathrm{disk}$ $\left[\frac{M_\ast}{100}\right]$ & $N_r$ & $N_\phi$ \\
				\hline
				\hline
				$-7$ & $-2$ & $1$ & $2.58$ & $435$ & $849$ \\
				$-7$ & $-2$ & $4$ & $2.57$ & $383$ & $748$ \\
				$-7$ & $-2$ & $10$ & $8.12$ & $441$ & $861$ \\\hline
				$-8$ & $-2$ & $1$ & $0.08$ & $699$ & $1364$ \\
				$-8$ & $-2$ & $4$ & $0.38$ & $651$ & $1271$ \\
				$-8$ & $-2$ & $10$ & $0.94$ & $527$ & $1029$ \\\hline
				$-8$ & $-3$ & $1$ & $0.49$ & $579$ & $1130$ \\
				$\bm{-8}$ & $\bm{-3}$ & $\bm{4}$ & $\bm{3.40}$ & $\bm{531}$ & $\bm{1037}$ \\
				$-8$ & $-3$ & $10$ & $9.10$ & $515$ & $1005$ \\\hline
				$-9$ & $-3$ & $1$ & $0.10$ & $813$ & $1587$ \\
				$-9$ & $-3$ & $4$ & $0.40$ & $685$ & $1337$ \\
				$-9$ & $-3$ & $10$ & $0.96$ & $529$ & $1033$ \\\hline
				$-9$ & $-4$ & $1$ & $0.77$ & $699$ & $1364$ \\
				$-9$ & $-4$ & $4$ & $3.84$ & $655$ & $1279$ \\
				$-9$ & $-4$ & $10$ & $9.51$ & $529$ & $1033$ \\\hline
			\end{tabular}
			\caption{Grid setup. Our fiducial model is shown in bold, and the same setup was used for runs with radiative diffusion.}
			\label{table-grid}
		\end{table}
				
		We first investigate the thermal input of a planet onto the disk and the structure of the gap that planet might possibly carve. To do so, we first present some comparisons of the gap width and depth across different models for both surface density and temperature and highlight the influence of disk aspect ratio and viscosity on the planet's ability to open a gap. We then select to show temperature profiles for various disks with identical initial temperatures, so that the planet's impact becomes more apparent.
			
		After that, we take a closer look at the structure of spiral arms by tracking the same quantities along their crests and compare them against the azimuthally averaged disk profiles. This would give us an estimation for both the temperature contrast between the spirals and the disk, as well as the shock strength along those spirals.
			
		\subsection{Gap opening capabilities of a planet}
		\label{section:results-profiles-gaps}
		
		\begin{figure*}[h]
			\centering
			\includegraphics[width=\textwidth]{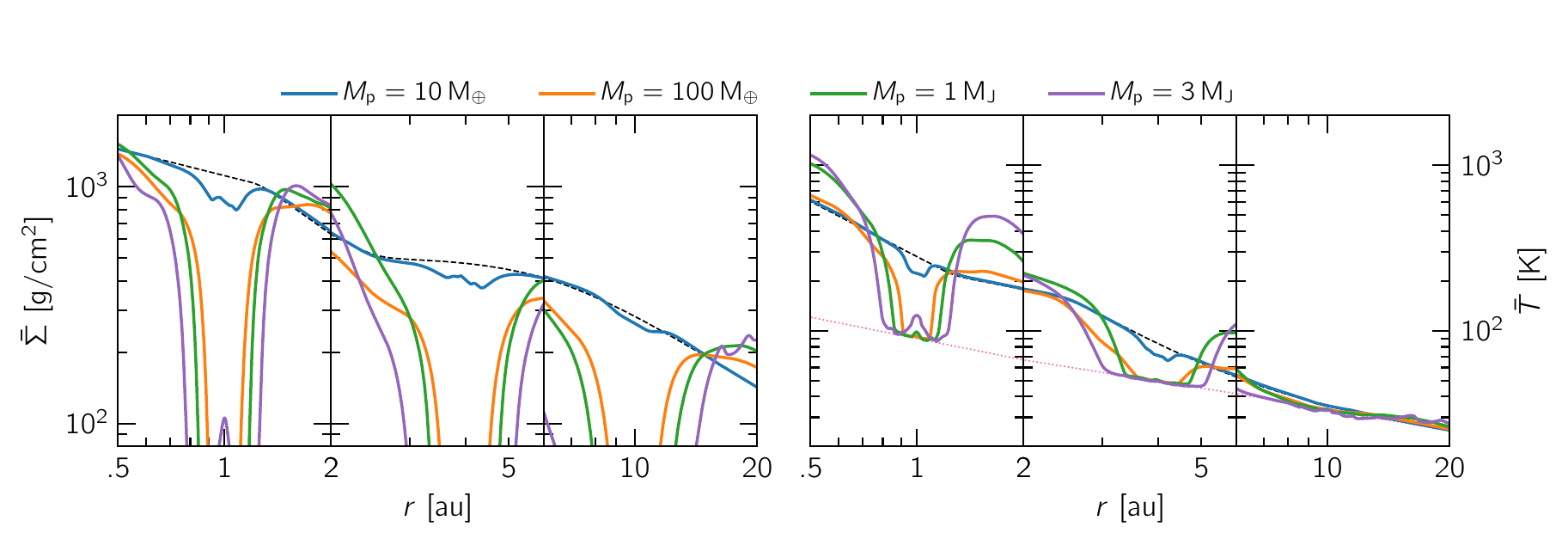}
			\caption{Azimuthally averaged profiles of surface density (left) and temperature (right) across planet masses and locations for our models. More massive planets open deeper and wider gaps, but the temperature inside the gap region is not necessarily lower in the outer, irradiation-dominated disk, due to stellar irradiation. The dotted pink line marks the disk irradiation temperature ($Q_\mathrm{cool}=Q_\mathrm{irr}$).}
			\label{fig:gaps-Mp}
		\end{figure*}
				
		\begin{figure*}[h]
			\centering
			\includegraphics[width=\textwidth]{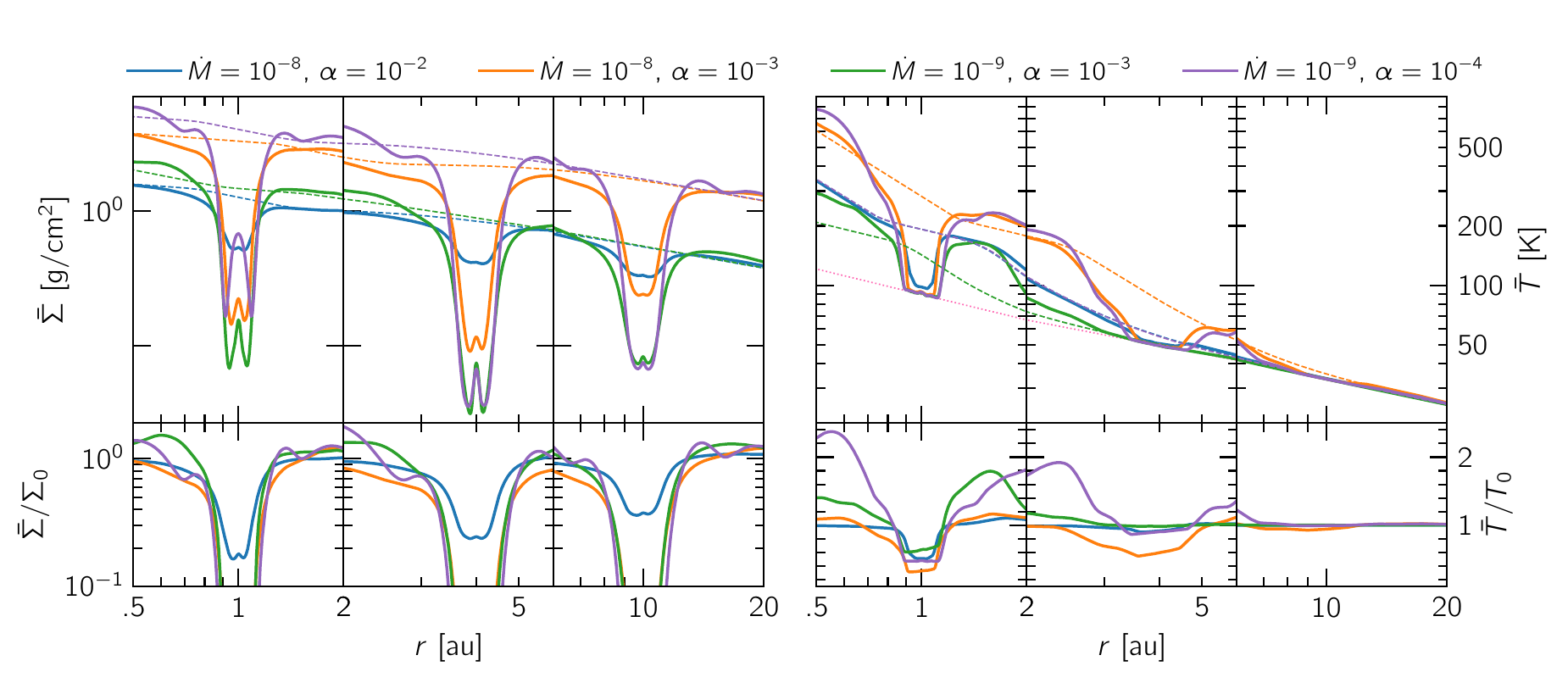}
			\caption{Azimuthally averaged profiles of surface density (left) and temperature (right) around the planet's location across different disk models for $M_\mathrm{p}=100$\,$\mathrm{M}_\oplus$. These snapshots are taken once each model has reached a quasi-equilibrium state, meaning that the gap depth is not well-defined for very low viscosities. However, it only takes a few hundred orbits for the overall disk structure to equilibrate.}
			\label{fig:gaps-rp}
		\end{figure*}
						
		On Fig.~\ref{fig:gaps-Mp} we compare the gap-opening capabilities of planets of different masses for our fiducial model ($\dot{M}=10^{-8}$, $\alpha=10^{-3}$). While the least massive planet in these models (10 $\mathrm{M}_\oplus$) does not open a gap, the rest are sufficiently massive to show a clear trend between planet mass and gap width, with more massive planets opening deeper and wider gaps.
			
		However, we also find that there is a lower limit to the temperatures inside the gap. This arises due to stellar irradiation, which provides enough heat to form an effective temperature floor where $Q_\mathrm{irr}=Q_\mathrm{cool}$. This term overpowers other heating effects with increasing radii and, as a result, a temperature gap is not visible in the outer disk regardless of planet mass.
		
		Then, for a given planet mass of $M_\mathrm{p}=100$\,$\mathrm{M}_\oplus$, we carry out the same comparison across models with different disk parameters. The results are shown on Fig.~\ref{fig:gaps-rp}, where a similar behavior is visible for temperatures inside the gap.
		
		 A key point is that we observe shallower gaps for higher values of $\alpha$ (for a given $\dot{M}$) or $\dot{M}$ (for a given $\alpha$). This can be understood by looking at the two main mechanisms determining the gap edge, as shown by \citet{crida-etal-2006}: viscosity and pressure gradients. Before adding a planet to a disk of a given $\dot{M}$ and $\alpha$, one can show that $\dot{M}\propto \nu\Sigma\propto \alpha p r^{3/2}$, such that pressure gradients are stronger in disks with a higher $\dot{M}$ or lower $\alpha$. This, combined with the fact that viscous dissipation scales with $\alpha$, allows for easier gap opening in disks with either a higher $\dot{M}$ or lower $\alpha$. This is nothing new, as it has been pointed out by \citet{crida-etal-2006}, \citet{zhang-etal-2018} and previous studies that disks with a lower viscosity or aspect ratio support the gap opening process.

	 	\begin{figure*}
	 		\includegraphics[width=\textwidth]{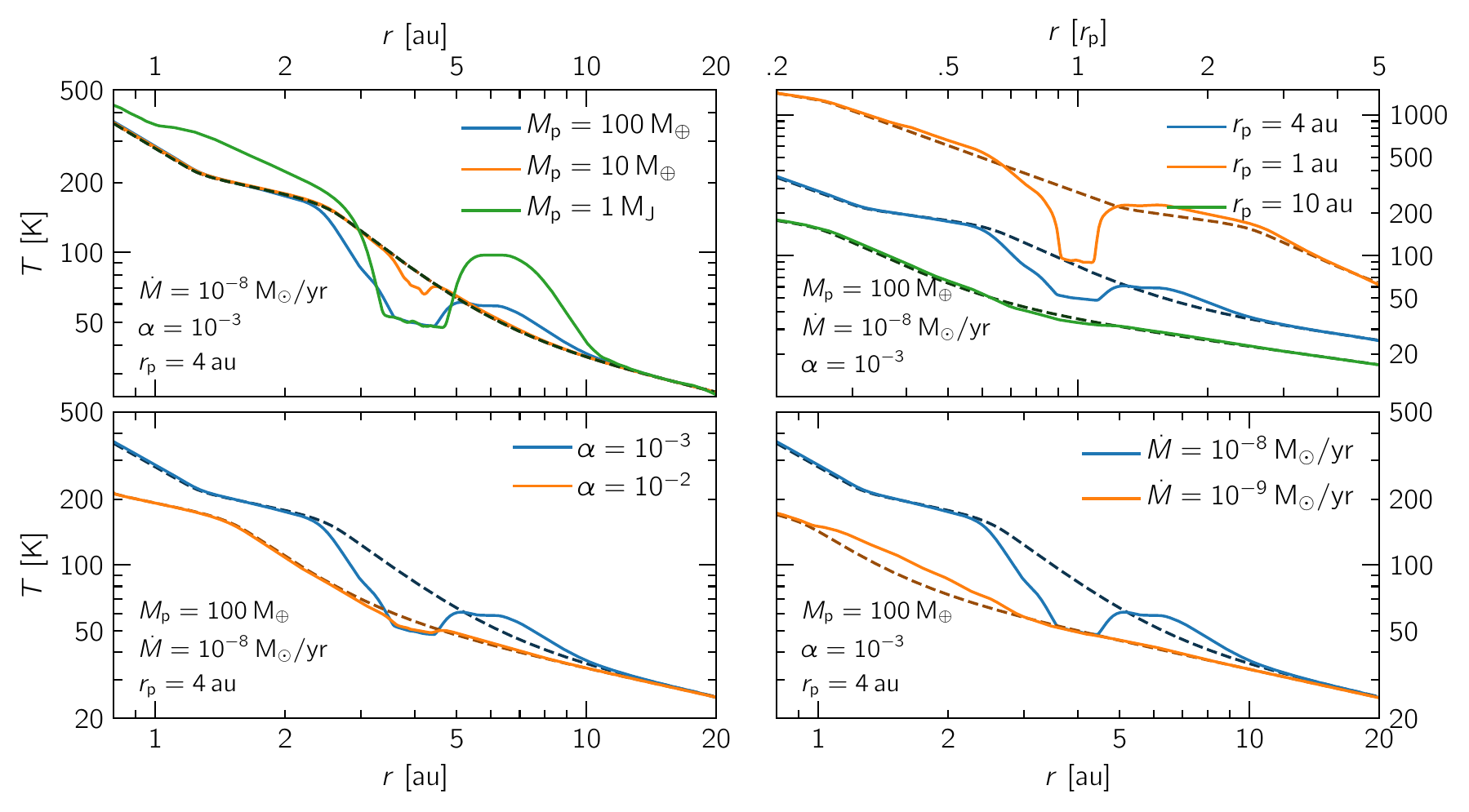}
	 		\caption{Azimuthally averaged temperature profiles for 4 pairs of models, showcasing the influence of each of the 4 variables in our parameter space ($M_\mathrm{p}$, $\dot{M}$, $\alpha$, $r_\mathrm{p}$) on spiral shock heating. Our fiducial model ($M_\mathrm{p}=100$\,$\mathrm{M}_\oplus$, $\dot{M}=10^{-8},\alpha=10^{-3},r_\mathrm{p}=4$) is shown in blue on every panel, while dashed lines depict initial profiles of the model of corresponding color. In all cases we notice stronger shock heating in optically thicker regions of the disks. This behavior can also be seen in the rest of our suite of models.}
	 		\label{fig:profiles}
	 	\end{figure*}
	 
	 	\subsection{Spiral shock heating by a planet}
	 	\label{section:results-profiles-shocks}
	 	
	 	On the right panels of Figs.~\ref{fig:gaps-Mp}~and~\ref{fig:gaps-rp}, apart of the depth (or lack of) a gap we notice a temperature increase on both sides of the planet's vicinity, sometimes by a factor of 1.5--2 compared to the initial profiles. This heat excess scales with planet mass, and can lead to quite high temperatures in the inner disk. This pattern is in agreement with our theoretical estimates in Sect.~\ref{section:shock-heating}, from which we should expect that the optically thick inner disk is more susceptible to heating by spiral shocks.
	 		
	 	In an attempt to compare the individual effect of each of our four parameters---planet mass, accretion rate, viscosity parameter and planet radius---we plot pairs of models where three out of four parameters are the same, allowing us to quantify the influence of the fourth. The trends we find with this method are clear enough that a comparison between our fiducial model and four other models suffices to convey the general picture. This comparison is shown on Fig.~\ref{fig:profiles}.
	 	
	 	\begin{figure}
	 		\includegraphics[width=.5\textwidth]{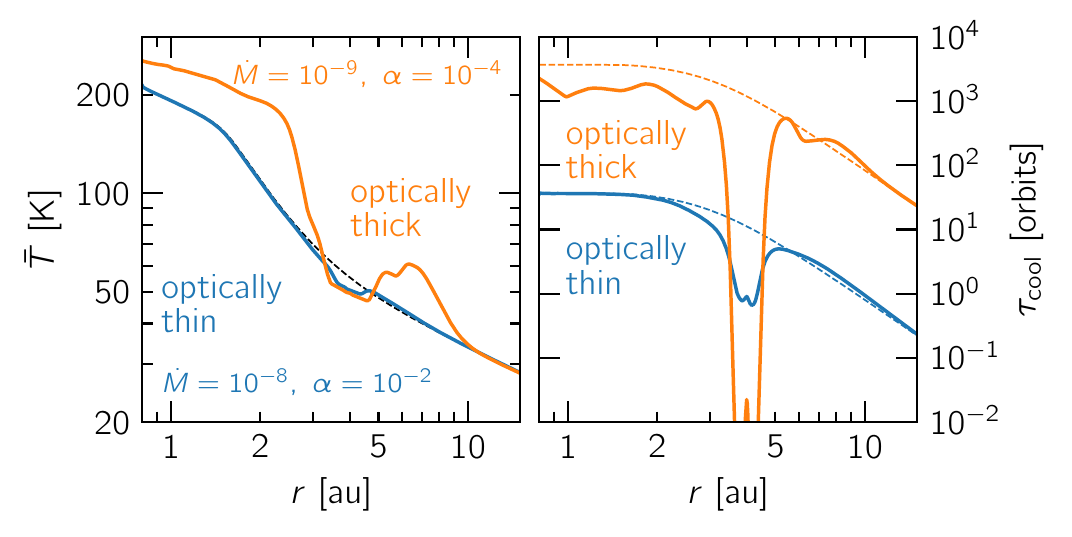}
	 		\caption{Azimuthally averaged temperature and cooling timescale for two models that share the same initial temperature profile, but have very different optical depth profiles. We see that shock heating is significantly stronger for the optically thicker model, while the optically thinner one looks almost unchanged.}
	 		\label{fig:timescale}
	 	\end{figure}
	 	
	 	The common denominator of these 4 panels is the cooling timescale of different disks and/or regions within them. We can get a rough estimate of this timescale by focusing on the cooling term in Eq.~\eqref{eq:source-terms} and writing
	 	\begin{equation}
	 		\label{eq:cooling-timescale}
	 		\frac{\partial\Sigma e}{\partial t}\sim \frac{\Sigma e}{\tau_\mathrm{cool}} \sim Q_\mathrm{cool} = \sigma_\mathrm{SB}\frac{T^4}{\tau_\mathrm{eff}} \Rightarrow \tau_\mathrm{cool} \approx \frac{\tau_\mathrm{eff}\mathrm{R}\Sigma }{\mu(\gamma-1)\sigma_\mathrm{SB} T^3},
	 	\end{equation}
	 	which further backs the assumption that the deciding factor in determining the contribution of shock heating to the thermal budget of the disk is the optical depth. For completion, we compare two models where $\{\dot{M}=10^{-8},\alpha=10^{-2}\}$ and $\{\dot{M}=10^{-9},\alpha=10^{-4}\}$ respectively. These two models happen to have identical initial temperature profiles, but show the lowest and highest optical depths in our suite of simulations, respectively. This comparison is plotted on Fig.~\ref{fig:timescale} and clearly shows the effect of optical depth on the contribution of shock heating. It should be noted that, even though the optically thinnest model on that Figure shows only small traces of excess heat due to shocks, the cooling timescale is still more than 10\% of the orbital period at 10\,au and therefore radiative effects of the disk should still be treated self-consistently to get a correct picture of its evolution.
	 	
	 	From our results, we can conclude that shock heating is in principle important for all of our models, and sometimes dominates when the cooling timescale of the disk is sufficiently long. This implies that planets with semimajor axes in the range of 1--10\,au can noticeably heat up their environment through spiral shocks and, as a result, an adiabatic equation of state is necessary when modeling planet--disk interaction in this regime.

		\begin{figure*}
			\includegraphics[width=\textwidth]{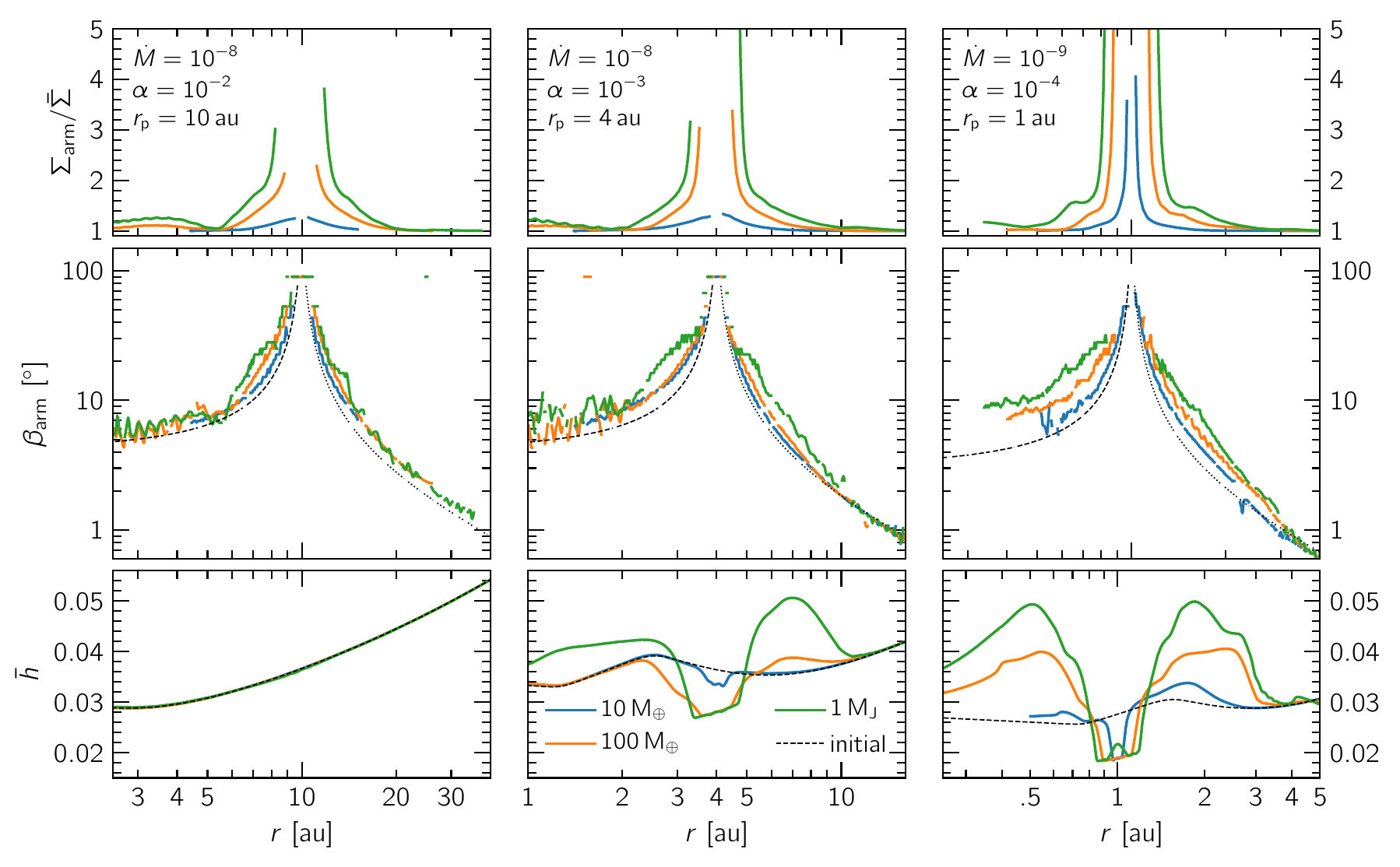}
			\caption{Shock strength, pitch angles and azimuthally averaged aspect ratios for 3 representative models. The optical depth and cooling timescale increase from left to right. Top: Shocks are stronger for more massive planets, but their dependence on disk parameters is not clear. To filter out unphysical results, we mask points that lie within the gap region ($|r-r_\mathrm{p}|\leq 2.5\,\mathrm{R_{Hill}}$) or the corotating region (see Eq.~\eqref{eq:horseshoe}). Middle: pitch angles are roughly the same regardless of planet mass for the optically thin case, but deviate with increasing optical depth. The black lines denote the expected values using Eq.~\eqref{eq:pitch-angles-ogilvie} for $h=h_\mathrm{p}$ (dashed) or Eq.~\eqref{eq:pitch-angles-muto} for $h=h_\mathrm{p} r^{2/7}$ (dotted). Bottom: A power law fit of the aspect ratio is only possible for the optically thin case or low mass planets, but the fidelity of such a fit breaks down even for 10-Earth-mass planets in an optically thick disk. The black dashed lines refer to initial aspect ratio profiles.}
			\label{fig:armstrength}
		\end{figure*}

		\subsection{Spiral arm structure and shock strength}
		\label{section:results-profiles-spirals}
		
		In the previous section we discussed the effect of shock heating by comparing azimuthally-averaged profiles in simulations with and without planets. While this is a useful approximation to form a general image of planet--disk interaction, it cannot isolate the contribution of individual spirals or their properties. Inspired by the approach of \citet{zhu-etal-2015}, we wrote a script that can trace spiral arms as they propagate away from the planet and log their coordinates as well as $\Sigma_\mathrm{arm}$ and $T_\mathrm{arm}$ along their crests. We then use this data to estimate a proxy for the shock strength along those spirals as $\Sigma_\mathrm{arm}/\bar{\Sigma}$ (shown in Sect.~\ref{section:shock-heating}), as well as their pitch angles $\beta$ defined as $\tan\beta=\mathrm{d}\log r_\mathrm{arm}/\mathrm{d}\phi_\mathrm{arm}$.

		As in the previous section, trends among models are clear enough such that we do not need to present results for our entire library of simulations. Instead, we take into account that the contribution of shock heating scales with the optical depth and show results for 3 regimes: the optically thinnest model ($\dot{M}=10^{-8},~\alpha=10^{-2},~r_\mathrm{p}=10$), the optically thickest one ($\dot{M}=10^{-9},~\alpha=10^{-4},~r_\mathrm{p}=1$), as well as our fiducial model ($\dot{M}=10^{-8},~\alpha=10^{-3},~r_\mathrm{p}=4$), which also happens to lie somewhere in the middle. For each model, we calculate the shock strength and pitch angles of primary spirals (i.e., those that connect to the planet). Since the pitch angle scales with $h$ far from the launching point according to linear theory, we also plot the azimuthally averaged aspect ratio $\bar{h}$.
		
		In an attempt to filter out unphysical shock strength values inside the low-density ring around the planet, we calculate the half-width of the horseshoe region as shown in \cite{paardekooper-etal-2010}:
		\begin{equation}
			\label{eq:horseshoe}
			x_\mathrm{h}=1.1r_\mathrm{p}\sqrt{\frac{1}{h_\mathrm{p}}\frac{M_\mathrm{p}}{M_\ast}}\left(\frac{0.4 H_\mathrm{p}}{\gamma \epsilon_\mathrm{p}}\right)^{1/4},
		\end{equation}
		as well as the shock length following \citet{zhu-etal-2015}:
		\begin{equation}
			\label{eq:shock-length}
			x_\mathrm{s} = 0.93 \left(\frac{\gamma+1}{12/5}\frac{M_\mathrm{p}}{M_\mathrm{th}}\right)^{-2/5}H, \quad M_\mathrm{th} 
			\equiv \frac{c_\mathrm{s}^3}{\mathrm{G}\Omega_\mathrm{p}},
		\end{equation}
		where $M_\mathrm{th}\approx 1 \mathrm{M_J}\left(\frac{h_\mathrm{p}}{0.1}\right)^3\left(\frac{M_\ast}{\mathrm{M_\odot}}\right)$ is the disk thermal mass. If $M_\mathrm{p}>M_\mathrm{th}$, then $x_\mathrm{s} = 0$ (spirals shock immediately upon launch).
		For more massive planets, where a gap opens, we set a cut-off where $|r-r_\mathrm{p}| \leq 2.5\,\mathrm{R_{Hill}}$. We then exclude data within any of those 3 regions.
		
		Our results are summarized in Fig.~\ref{fig:armstrength}. We see that the shock strength of spirals typically lies between 1.5--3 for massive planets, but rarely exceeds 1.5 for 10\,$\mathrm{M}_\oplus$ models. Looking at Fig.~\ref{fig:heatingrates}, we can see that such shocks produce competitive heating when compared to either viscosity or stellar irradiation for $r_\mathrm{p}\leq4$. In the $r_\mathrm{p}=10$ case (left panels of Fig.~\ref{fig:armstrength}), however, the planet is embedded in an optically thin, irradiation dominated region and as a result shock heating is overcome by stellar irradiation, which eventually sets the overall profile.
		
		As far as pitch angles are concerned, we attempt to fit their curves with analytical formulas that assume an aspect ratio profile. In the inner disk, due to the temperature being defined by different power laws depending on the opacity regime, it's easier to assume that the aspect ratio is roughly constant and use the formula by \citet{ogilvie-lubow-2002} to calculate the location of spirals
		\begin{equation}
			\label{eq:pitch-angles-ogilvie}
			\phi_\mathrm{arm} = -\mathrm{sgn}(r-r_\mathrm{p}) \frac{2}{3h}\left[\left(\frac{r}{r_\mathrm{p}}\right)^{3/2}-\frac{3}{2}\ln \left(\frac{r}{r_\mathrm{p}}\right)-1\right].
		\end{equation}
		On the other hand, in the irradiation dominated outer disk (where $Q_\mathrm{irr}\approxeq Q_\mathrm{cool}$), we can utilize the formula by \citet{muto-etal-2012}:
		\begin{equation}
			\label{eq:pitch-angles-muto}
			\begin{split}
			\phi_\mathrm{arm} &= -\frac{\mathrm{sgn}(r-r_\mathrm{p})}{h_\mathrm{p}}\\
			&\phantom{=}\times \left[ \left(\frac{r}{r_\mathrm{p}}\right)^{1+\eta}\left\{\frac{1}{1+\eta}-\frac{1}{1-\zeta+\eta}\left(\frac{r}{r_\mathrm{p}}\right)^{-\zeta}\right\}\right.\\
			&\phantom{=}\left.-\left(\frac{1}{1+\eta}-\frac{1}{1-\zeta+\eta}\right)\right],
			\end{split}
		\end{equation}
		where $\Omega_\mathrm{K}\propto r^{-\zeta}$ and $h\propto r^{0.5-\eta}$. For $h\propto r^{2/7}$ (see Eq.~\eqref{eq:source-terms}), we have $\zeta=\nicefrac{3}{2}$ and $\eta=\nicefrac{3}{14}$.

		Even by utilizing both formulas, we find that it is difficult to get a good fit and that we always underestimate the pitch angles. This shows that the heating generated by the planets' spirals can change the disk aspect ratio such that it cannot be accurately approximated with a power law. The fit completely breaks down for massive planets or optically thick disks, where we find that pitch angles are inflated around the location of planets. This makes sense, since shock heating peaks at these locations (e.g., see Figs.~\ref{fig:heatingrates}~and~\ref{fig:gaps-rp}).

		\begin{figure*}[h]
			\centering
			\includegraphics[width=.9\textwidth]{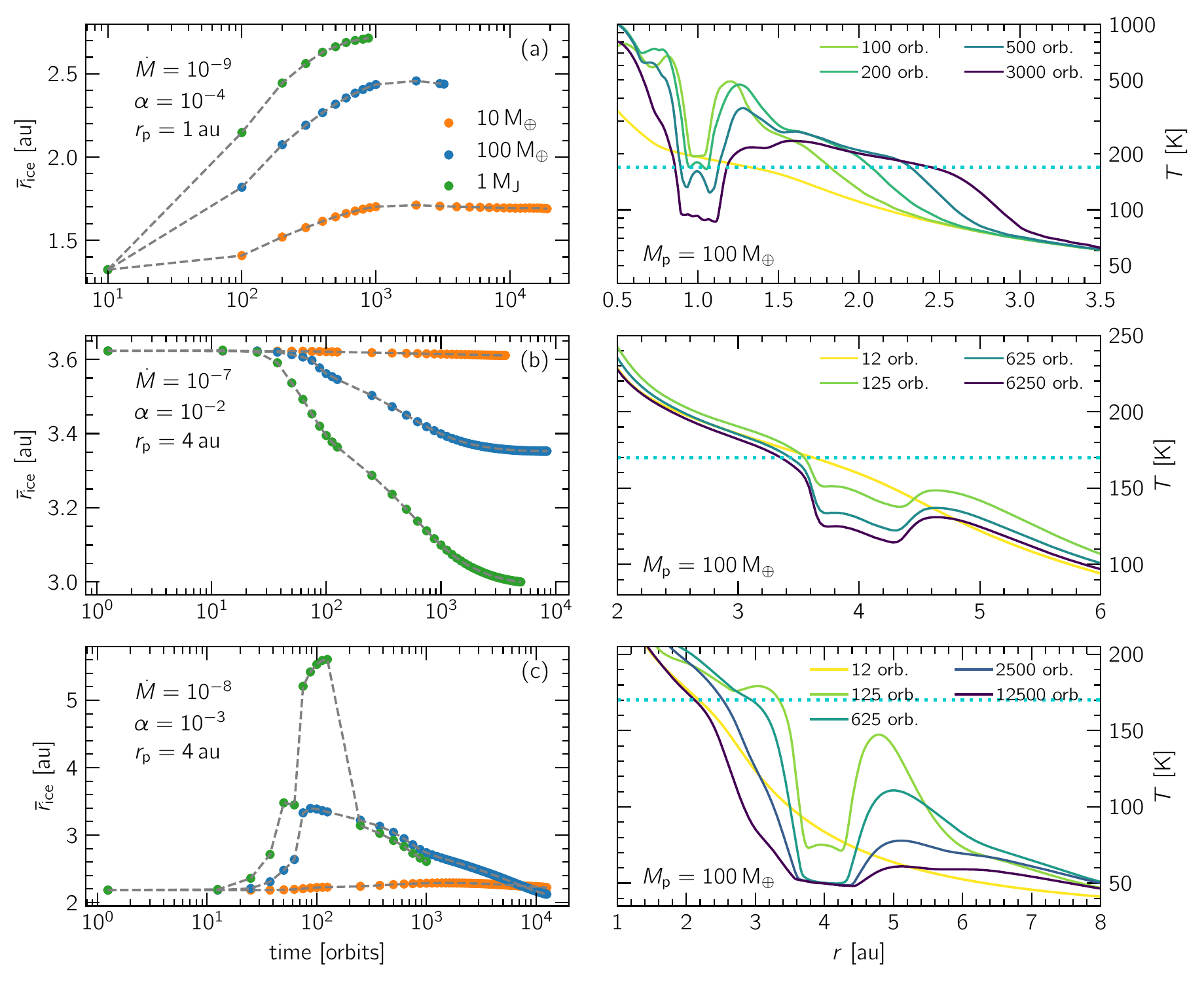}
			
			\caption{Azimuthally-averaged iceline location for 3 sample models, showcasing the 3 possible evolution scenarios. (a): Spiral heating pushes the iceline outwards. (b): The iceline recedes to the inner gap edge. (c): Shock heating initially pushes the iceline outwards, but eventually a gap is carved and the iceline recedes inwards. These effects are amplified for more massive planets. It should be noted that, in the case of a cold gap, ice can recondense within the gap region (panel~(a)).
			}
			\label{fig:icelines}
		\end{figure*}
			
	\begin{figure*}[h]
		\centering
		\includegraphics[width=\textwidth]{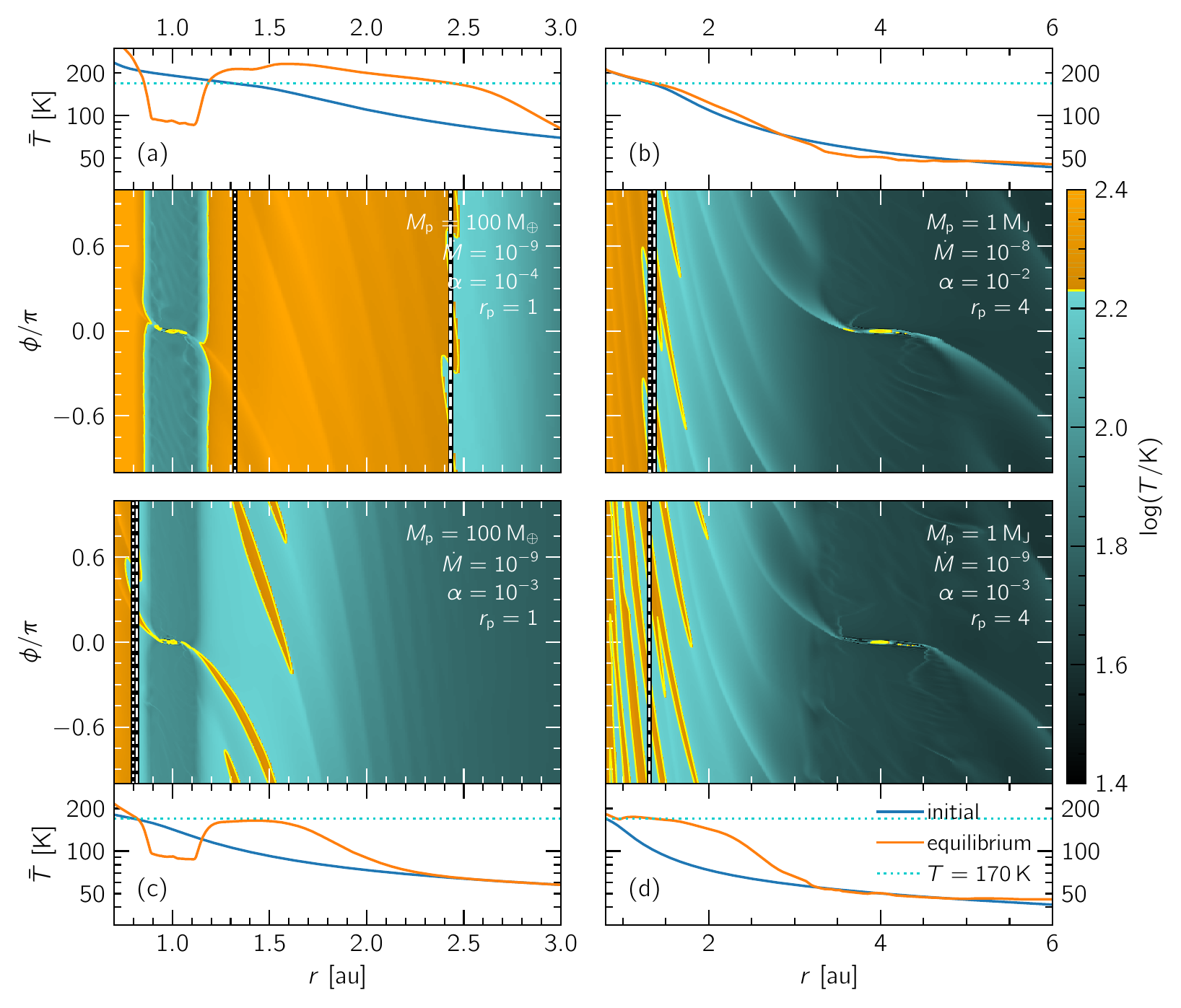}
		\caption{Temperature maps for 4 models, showcasing the azimuthal structure of the iceline. The two colormaps denote the blue, "cold" ($T<T_\mathrm{ice}$) and orange, "hot" ($T> T_\mathrm{ice}$) disk, respectively, with a yellow line separating the 2 regions ($T=T_\mathrm{ice}$). The initial location of the iceline (before a planet is embedded) is marked with a dotted white line, while its azimuthally averaged location in equilibrium is shown with a dashed white line.}
		\label{fig:slushies}
	\end{figure*}

		\section{Location and shape of the water iceline}
		\label{section:icelines}
		
		Using the analytical estimates in Sect.~\ref{section:shock-heating}, we showed that if a massive planet is located in the optically thick part of the disk the shocks are adiabatic, and spiral heating can increase the temperature of the disk. In Sect.~\ref{section:profiles}, we presented results on the heating potential of these shocks through our numerical simulations, confirming our estimates. In this section, we investigate \emph{how much a planet can displace icelines (e.g., the water iceline) in a disk.} Our motivation to do so lies in quantifying the possibility that a planet can starve itself or the inner disk of water as it forms, as well as its ability to change the environment in which planetesimals could grow \citep{drazkowska-alibert-2017}.

		From Fig.~\ref{fig:opacity} it becomes clear that the Rosseland mean opacity is independent of density up to $T\approx10^3$\,K, such that the water iceline effectively represents the temperature where ice sublimates. This implies that the location of the iceline does not explicitly depend on the density jump across shocks, but instead on the temperature they can reach. In this study, we follow \citet{lin-papaloizou-1985} and define the water iceline $r_\mathrm{ice}$ as the point where the temperature reaches $T_\mathrm{ice}=170$\,K (see opacity transition at this point in Fig.~\ref{fig:opacity}).

		\subsection{Location of the azimuthally-averaged water iceline}
		\label{section:results-aziavg}

		In the absence of a planet, and assuming an axisymmetric disk, the equation $r=r_\mathrm{ice}$ defines a circle with radius $r_\mathrm{ice}$ from the star, within which water can only be found in the form of vapor. For now, let us assume that the presence of the planet does not significantly perturb the iceline in the azimuthal direction but instead moves it uniformly towards or away from the star, such that the new location of the iceline is $\bar{r}_\mathrm{ice}=\langle r_\mathrm{ice}\rangle_\phi$ once equilibrium is reached.
		 
		Depending on its initial location, the iceline is susceptible to two planet-induced phenomena: the closer it lies next to the planet, the more it is exposed to shock heating from the latter, leading to a larger outward displacement as the planet heats up its surroundings. In the extreme case that the iceline initially lies directly next to the planet, it can be pushed away by a factor of 1.5 or even more (see Fig.~\ref{fig:icelines}a). The impact on the location of the iceline scales with the mass of the planet, with Jupiter-sized planets increasing $\bar{r}_\mathrm{ice}$ by a factor of 2.
		
		However, in the case that the planet is massive enough to open a gap, embedding it too close to the iceline such that the latter overlaps with the optically thin gap region results in a recession of the iceline towards the inner gap edge (see Fig.~\ref{fig:icelines}b). In this case, the final location of the iceline depends on the width of the gap, which also scales with planet mass. Of course, for planets of sufficiently low mass a gap will not open and therefore the iceline location will essentially remain intact. 
		
		Finally, it is possible that gap opening and shock heating can compete for the determination of the location of the iceline. This behavior is shown on Fig.~\ref{fig:icelines}c: shock heating initially moves the iceline outwards, "pulling" it closer to the planet. However, the slower gap opening process eventually "catches up" and the steep temperature gradient near the inner gap edge extends the region affected by the gap down to around $2.5$\,au ($0.6$\,$r_\mathrm{p}$), returning the iceline to a location similar to its initial one once the gap has fully opened.
		
		\subsection{Azimuthal structure of the iceline}
		\label{section:results-azi}

		In the previous section we considered the iceline to be an axisymmetric line---a circle with radius $\bar{r}_\mathrm{ice}$ around the star---by tracking its location using azimuthally-averaged temperature profiles. However, shock heating is a strongly non-axisymmetric process as it follows the trajectories of spiral arms. Therefore, its influence on the location of the iceline should introduce azimuthal features on the latter.
		In other words, the full picture is 2D---at least within the scope of this project.
				
		To examine the azimuthal structure of the iceline, we plot temperature maps of our models at equilibrium and draw contours at $T(r,\phi)=T_\mathrm{ice}$. We summarize the results for some of our representative models in Fig.~\ref{fig:slushies}.
			
		In general, a few key behaviors can be observed in our simulation results. The iceline tends to move outwards in optically thick disks (as shown above), and deform such that it follows the trajectories of spirals when strong shocks are present. Therefore, we can distinguish a few distinct ``extremes''. For a low-mass planet in an optically thin disk, the location or shape of the iceline do not change. In an optically thick disk, the same planet might slightly move the iceline outwards or perturb it along the azimuthal axis.
		
		In both of these cases, the analysis in Sect.~\ref{section:results-aziavg} still applies with good accuracy. However, a massive planet launches strong shocks and opens a gap, which can halt the outward movement of the iceline or even cause it to recede to the inner gap edge. Therefore, for a high-mass planet in an optically thin disk (where typically $r_\mathrm{ice}^{t=0}<r_\mathrm{p}$), we see hot spirals form in the \emph{inner} disk (such that $T_\mathrm{arm}>T_\mathrm{ice}$) but little to no radial displacement of the iceline (see Fig.~\ref{fig:slushies}\,b). If initially $r_\mathrm{ice}^{t=0}\sim r_\mathrm{p}$, the iceline will recede to the inner gap edge in addition to forming hot spirals in the \emph{outer} disk (see Fig.~\ref{fig:slushies}\,c).
		
		On the other hand, for a massive planet in an optically thick disk, shock heating is strong enough to displace the iceline to the outer disk, to the point where spiral pitch angles are small and the spirals are very tightly wound, heating the disk uniformly in azimuth. In this case, the domain is split into a hot inner disk, a cold gap, and a hot ring in the outer disk (see Fig.~\ref{fig:slushies}\,a). If initially $r_\mathrm{ice}^{t=0}\sim r_\mathrm{p}$, the iceline will again recede to the inner gap edge while possibly forming hot spirals or a hot ring in the outer disk, depending on the optical depth.
		
		Of course, if $r_\mathrm{ice}^{t=0}\gg r_\mathrm{p}$, far out at the irradiation dominated outer disk, the iceline will not change in shape or location but a cold ring can still form inside the gap region. However, the optical depth rapidly increases at small distances from the star, and as such the pile-up of inner spiral arms by a Jupiter-sized planet  can still cause substantial heating, moving the iceline outwards even in optically thinner models (as shown in Fig.~\ref{fig:icelines}\,d).

	\section{Discussion}
	\label{section:discussion}
	In this section we discuss our findings with respect to their possible impact on the growth and change of orbital elements of the planets, and the structure of the disk. 
	\subsection{A shift of the iceline}
	Under certain conditions (low $\dot{M}$, low $\alpha$), a planet located at 1\,au could push the iceline outwards by a few au.
	This leads to a reduction of icy solid material present in the disk in size and number. This lowers the accretion of solid material
	onto the planet as more matter is in gaseous form which has a lower (viscous) drift than embedded particles.
	On the other hand, when the snow line moves just a little beyond the planet, icy aggregates that may fall apart can release tiny silicate grains
	\citep{schoonenberg-etal-2018}, and tiny dust may be less well trapped in the outer edge of the planet gap,
	possibly affecting accretion of dust onto the planet. The net effect will be an enhancement of dry over wet particle accretion onto the planet,
	and a reduction of the water/ice content in the inner regions of the disk.
	\subsection{Slush islands}
	In our simulations we found regions where the conditions within the disk are such that the temperature along the spirals is above the ice sublimation threshold and drops below it between spiral crests (see Fig.~\ref{fig:slushies}).
	Ice sublimates around the peak of the shock but condenses again further away from it, such that along the boundary of the spiral
	(as defined by the iceline) one might find a mixture of ice and water vapor with a "slushy" consistency,
	hence, we call them \emph{slush islands}.
	These may occur inside as well as outside of the planet's location.
	The repetitive sublimation/condensation will slow down dust growth as growing particles will periodically be reduced in size. 
	\subsection{Migration torques}
	The disk heating of an embedded planet will change the torques acting on it and hence its migration rate. As the torques scale inversely with the
	disk's scale height \citep{kley-nelson-2012}, it is expected that the planets slow down due to the heating they produce. This possibility was already explored and discussed by \citet{hallam-paardekooper-2018}, who showed that even a simplistic prescription of gap edge illumination can result in slowing down or even reversing the migration rate of planets. In light of our results concerning the planets' ability to heat up their vicinity through spiral shocks, our modeling supports the findings of that study in that regard.
	\subsection{Temperature within the gap}
	In our simulations the temperature within the gap was lower that the environment because of the reduced optical depth. For deep gaps the irradiation
	temperature was reached. Previous works have looked at the gas temperature in a planet’s gap (via radiative transfer calculations) considering the three dimensional vertical extent of the disk \citep{jang-condell-turner-2012}. In that study, the gap's temperature is determined by shadowing and illumination effects,
	which are not included in our 2D treatment of irradiation. It was also found that the outer gap edge, which is directly exposed to starlight, can heat
	the interior of the gap such that the temperature can be even higher than the ambient temperature.
	
	\subsection{Simplifications and assumptions}

	Throughout our study, we have made several assumptions about the various physical processes at play. We therefore find it important that we bring attention to the potential impact they can have on our results.
	
	First and foremost is our two-dimensional approximation in simulating global, adiabatic disks. \citet{lyra-etal-2016} pointed out that the additional degree of freedom in the vertical expansion of an adiabatic shock results in overall weaker shocks, suggesting that our results overestimate shock heating. This can be amended by ``scaling'' our results to refer to more massive planets.
		
	Secondly, regarding the smoothing length chosen for the planet's gravitational potential, we chose to evaluate the scale height locally ($H(r,\phi)$) instead of using that at the planet's location ($H_\mathrm{p}$). The reason behind this choice is that it corrects for the disk's finite thickness, as shown by \citet{mueller-etal-2012}. However, this assumption might be dangerous in radiative simulations. For example, the planet's accretion luminosity can result in a ``hot bubble'' around the planet \citep{klahr-kley-2006}, where the scale height can increase sharply with respect to its initial value. Nevertheless, this smoothing length becomes important at a scale far smaller than the planet's Hill radius and therefore should have a minuscule effect on the latter's gravitational potential.
		
	Additionally, our model of stellar irradiation contains a simplification in that disk self-shadowing is ignored. Specifically, we assume that the star illuminates a disk where the scale height does not significantly change (such that $\D{\log H}{\log r}$ is constant and refers to a power-law profile for $H$), but then point out that shock heating \emph{can} in fact strongly affect said disk property. While this assumption leads to a very straightforward and stable numerical implementation of stellar irradiation, it occasionally results in a disparity between our assumption of $\nicefrac{9}{7}$ for $\D{\log H}{\log r}$ and the actual value. This disparity is greater for optically thick disks, and vanishes with increasing distance from the star. We can therefore justify our choice by remembering that stellar irradiation is indeed a dominant heat source at large radii, where the approximation holds best, and gives way to viscous/shock heating near the star, rendering it insignificant regardless of how well the approximation holds.

	\section{Conclusions}
	\label{section:conclusions}
	
	We examined the thermodynamical impact of planets on the ambient protoplanetary disk in which they are embedded. To do so, we first calculated an estimate for the amount of heat a planet can deliver into the disk through spiral shocks and showed that such heating can be significant. We noted that this process is strongest at the immediate vicinity of the planet, but has the potential to influence a larger area depending on disk optical depth. We then carried out a grid of 2D numerical hydrodynamics simulations with included radiative effects in order to find out if and how much this planet-generated heating can influence the disk and displace or deform the otherwise axisymmetric water iceline, defined as the radius $r_\mathrm{ice}$ where $T(r_\mathrm{ice})=T_\mathrm{ice}=170$\,K.
	
	We found that spiral shock heating is most important in optically thick, viscosity-dominated disks. Both of these requirements suggest a long cooling timescale, and are fulfilled in the inner few au of a protoplanetary disk. On the other hand, the irradiation-dominated outer disk suppresses shock heating by raising the aspect ratio. However, even when a planet is embedded in the outer disk, its inner spirals can heat up the disk as they propagate inwards.
		
	We also showed that both a high viscosity or aspect ratio inhibit the gap opening process, a result which is consistent with previous studies. On top of that, treating radiative effects allows us to probe the gas temperature inside the gap region. We found that in the inner disk, where viscous heating determines the gas temperature, a cold gap can be seen around massive planets. This is not visible in the outer disk, where the temperature both inside and out of the gap region is determined by the irradiation temperature.
	
	By tracing the planet's spiral arms we found that planet-induced spiral shocks scale in strength with planet mass, such that shock heating is strongest for massive planets. This led to a noticeable difference between spiral and background temperatures, with clear implications on the pitch angles of said spirals. We also showed that, due to the fact the aspect ratio can increase dramatically by high-mass planets, fitting pitch angles with a standard flaring-aspect-ratio prescription will in principle not yield accurate results when shock heating is important.
	
	We then investigated the planet's ability to displace the water iceline. We found that shock heating by the planet can increase temperatures enough to push the iceline away from the star. This outward displacement of the iceline can happen with various degree, depending on the optical depth of the disk. Optically thicker disks are unable to efficiently radiate away excess heat, and are prone to larger iceline displacements. Shock heating can then lead to either a uniform outward movement and/or a non-axisymmetric deformation of the iceline.
	
	In the inner few au of our disks, planets that are massive enough to carve a gap can create a cold ring around their semimajor axis. This gap cooling effect can easily overpower shock heating in the immediate vicinity of the planet, pulling the iceline inwards to the inner gap edge if it was initially near the soon-to-be-opened gap region, or creating a ``hot ring'' outside of the planet's location if the iceline maintains a radius greater than that of the planet's semimajor axis.
	
	However, it is also possible that the iceline deforms due to the temperature contrast between spirals and the disk background, such that it bends to follow spiral trajectories. This deformation is clearest for strong shocks in optically thin disks, where the iceline can trace the inner or outer spirals depending on the initial disk temperature profile. Such spirals will then be water-poor, with possible implications on dust growth in their vicinity. These effects can impact the accretion rate and composition of accreted particles on a planet.
	
	As far as the scale of this displacement and/or deformation is concerned, we find that planet mass plays a leading role in determining both shock strength as well as gap width, such that any effect related to the location or shape of the iceline is amplified for higher planet masses.

	Finally, we report that accounting for radiative diffusion in the disk midplane leads to no significant differences in temperature profiles or iceline deformation, as well as barely any observable differences in azimuthally-averaged iceline locations and spiral arm opening angles. As a result, it can be safely ignored in this context.
	\begin{acknowledgements}
		The authors thank the referee for the constructive comments and suggestions towards improving the quality of the manuscript.
		
		SA acknowledges support from the Swiss NCCR PlanetS.
		
		 CPD and WK acknowledge funding from the DFG research group FOR 2634 ``Planet Formation Witnesses and Probes: Transition Disks'' under grant DU 414/22-1 and KL 650/30-1.
		We acknowledge the support of the DFG priority program SPP 1992 "Exploring the Diversity of Extrasolar Planets under grant KL 650/27-1".  The authors acknowledge support by the High Performance and
		Cloud Computing Group at the Zentrum f\"ur Datenverarbeitung of the University
		of T\"ubingen, the state of Baden--W\"urttemberg through bwHPC and the German
		Research Foundation (DFG) through grant no INST 37/935-1 FUGG.
		
		This research was supported by the Munich Institute for Astro- and Particle Physics (MIAPP) of the DFG cluster of excellence ``Origin and Structure of the Universe".
		
		All plots in this paper were made with the Python library \texttt{matplotlib} \citep{hunter-2007}.
	\end{acknowledgements}

\bibliographystyle{aa}
\bibliography{refs}

\begin{appendix}

\section{Grid and code validation and physics justification}
\label{appendix:verification}

\begin{figure*}[h]
	\centering
	\includegraphics[width=\textwidth]{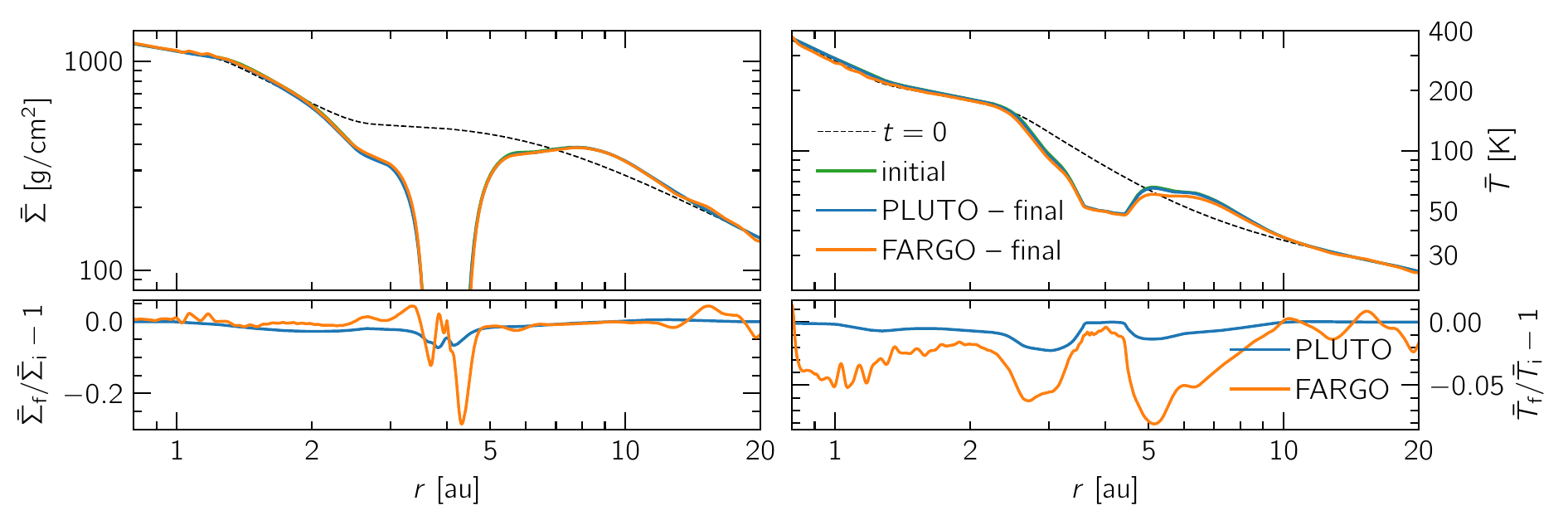}
	\caption{Comparison between \texttt{PLUTO} and \texttt{FARGO}, after restarting from an identical disk state in quasi-equilibrium ($t_\mathrm{i}=6\,250$~orbits) and running independently until $t_\mathrm{f}=t_\mathrm{i}+650$~orbits. Both surface density and temperature profiles are in good agreement across codes in the outer disk, while the different treatment of shock heating between the codes becomes evident only near the planet.}
	\label{fig:pluto-vs-fargo}
\end{figure*}

In order to verify our setup, we run a comparison test against the numerical hydrodynamics  code \texttt{FARGO} \citep{masset-2000}. For this test, we simulate the first 6\,250 orbits (50\,kyr) for a fiducial model ($M_\mathrm{p}=100$\,$\mathrm{M}_\oplus$, $r_\mathrm{p}=4$\,$\mathrm{au}$, $\dot{M}=10^{-8}$\,$\mathrm{M_\odot/yr}$, $\alpha=10^{-3}$) using \texttt{PLUTO}, then transfer the current, quasi-equilibrium disk state to \texttt{FARGO} and run for an additional 650~orbits using both codes using exactly the same physics. The final disk state for surface density and temperature is plotted in Fig.~\ref{fig:pluto-vs-fargo}. We find that the two codes produce similar results in the inner and outer disk, and differ only around the planet. We rationalize this by pointing out the fundamentally different treatment of shocks between the two codes:
\texttt{PLUTO} utilizes a Godunov-type scheme (a conservative, finite-volume approximation combined with a Riemann solver) that captures and resolves shocks with very good accuracy, in contrast to \texttt{FARGO}'s treatment of shocks.

Overall, the level of agreement between the two codes provides good grounds that our setup is working as intended, and that we can proceed with simulating our suite of models using \texttt{PLUTO}.

\begin{figure}[h]
	\centering
	\includegraphics[width=.5\textwidth]{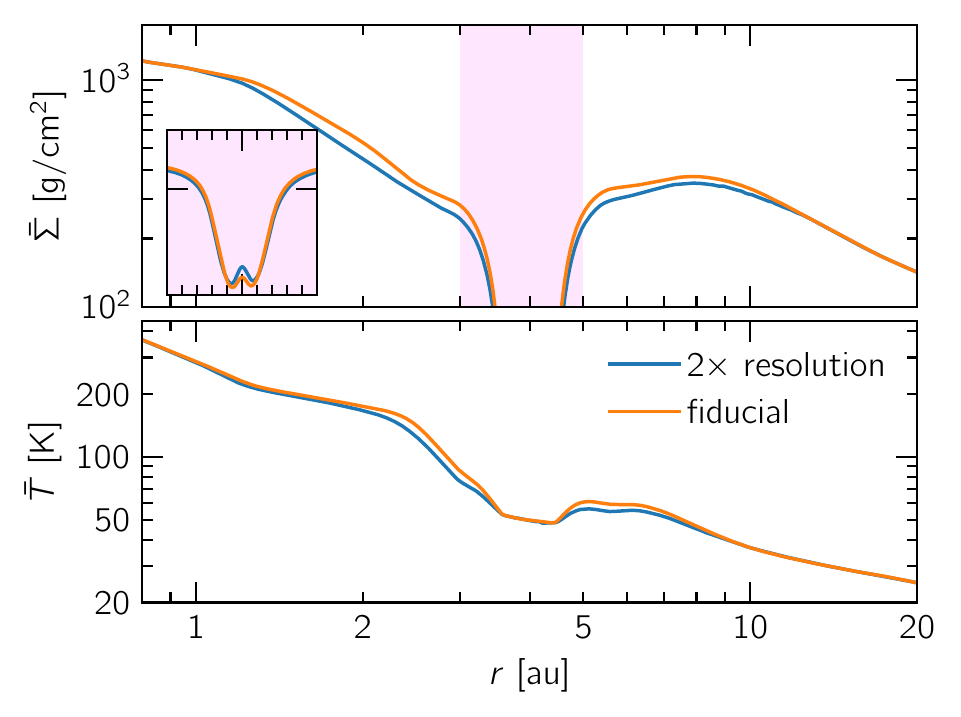}
	\caption{Two runs for our fiducial model, one of which (blue curves) has double the resolution on both the $r$ and $\phi$ axes (for a total of 4 times as many cells). The snapshots are taken at $t=100$\,kyr (12\,500 orbits), where equilibrium has more or less been reached. The inset zooms in on the pink-tinted region, showcasing the match between the two gap profiles.}
	\label{fig:high-res}
\end{figure}

Next, we verify our grid size. We use enough cells in the radial direction such that the pressure scale height $H$ is resolved by 6 or more cells at the planet's location, and an appropriate grid size in the azimuthal direction to maintain square cells (roughly twice the number of radial cells). To check whether this grid size is large enough, we rerun our fiducial model with double the resolution on both the $r$ and $\phi$ axes (using \texttt{PLUTO}) and compare the azimuthally-averaged surface density and temperature profiles (see Fig.~\ref{fig:high-res}). The results are quite similar (to roughly 90\%), so for our qualitative study this resolution of 6 cells per scale height is justified. The grid size used for our simulations is shown in Table~\ref{table-grid}.

Finally, we investigate the influence of radiative diffusion on the phenomena we would like to study, namely the shock strength of planetary spirals and the influence of the planet's shock heating on the water iceline. We find that accounting for radiative diffusion within the disk midplane barely affects the outcome of the two simulations. It was enabled in the fiducial model for $M_\mathrm{p}=10$ and $100$\,$\mathrm{M_\oplus}$, such that a case where no gap opens can also be studied. We report on the effect of radiative diffusion for these two models in more detail in Appendix~\ref{appendix:fld}.

\section{Grid structure}
\label{appendix:grid-structure}

As described in Sect.~\ref{section:numerics}, our first step is to generate 1D models for various combinations of $\dot{M}$ and $\alpha$. These models are calculated for $r\in [0.2,100]$\,au and then an appropriate region is selected depending on planet semimajor axis $r_\mathrm{p}$ by constructing a grid that mimics the \texttt{PLUTO} grid structure and fitting our initial profiles onto it through linear interpolation. That grid typically extends from $r_\mathrm{p}/5$--$5r_\mathrm{p}$ except for simulations with $10$\,$\mathrm{M_\oplus}$ planets, which were carried out earlier with a domain always between 0.5--20\,au regardless of planet location. A verification test was carried out to make sure that that setup did not affect the quality of the simulations and produced results identical to those using the former setup, therefore these simulations did not need to be rerun.

As far as grid size is concerned, we measure the pressure scale height $H_\mathrm{p}$ at the planet's location and construct a logarithmically-spaced array of $N_r$ cells in the $r$-direction that satisfies:
\begin{equation}
	\frac{H_\mathrm{p}}{\mathrm{\Delta}r_\mathrm{p}}\geq 6,
	\quad
	H_\mathrm{p} = \left.
	\frac{{c_\mathrm{s}}_\mathrm{iso}}{\Omega_\mathrm{K}}
	\right\vert_\mathrm{p,t=0}
	=\sqrt{\frac{RT_\mathrm{p}r_\mathrm{p}^3}{\mu \mathrm{G}M}},
	\quad
	M=M_\ast+M_\mathrm{p},
\end{equation}
while the number of cells in the azimuthal direction is chosen so that cells are square, or:
\begin{equation}
	N_\phi = \left|\left|\pi\frac{a+1}{a-1}\right|\right|,
	\quad
	a= \left(\frac{r_\mathrm{out}}{r_\mathrm{in}}\right)^{1/N_r}
\end{equation}
This typically results in grids of around $600\times1200$ cells (e.g., our fiducial model with $M_\mathrm{p}=100$\,$\mathrm{M_\oplus}$, $\dot{M}=10^{-8}$\,$\mathrm{M_\odot/yr}$, $\alpha=10^{-3}$, $r_\mathrm{p}=4$\,au contained $531\times1037$ cells).

\section{Effect of radiative diffusion}
\label{appendix:fld}

We repeat two simulations for our fiducial model ($r_\mathrm{p}=4$\,$\mathrm{au}$, $\dot{M}=10^{-8}$\,$\mathrm{M_\odot/yr}$, $\alpha=10^{-3}$) with flux-limited diffusion (FLD) enabled in the disk midplane. This additional term should smear out peaks in the temperature structure of the disk and could become important along the trajectories of spirals. We choose $M_\mathrm{p}\in \{10,100\}$\,$\mathrm{M_\oplus}$ to examine its overall impact on the disk whether a gap is carved or not. However, neither in the low- nor in the high-mass simulations do we see a significant difference, except for the lower peak temperature of spirals in the FLD models and a slight inward movement of the iceline. This implies that vertical cooling happens at a much faster rate than the planar diffusion timescale. By comparing their timescales, we indeed find that thermal cooling readjusts disk temperatures roughly 100 times faster than radiative diffusion does (except for inside the gap region), such that its effect on temperature profiles and the iceline is negligible.
	
	\subsection*{Disk profiles and spiral arms}
	\label{appendix:fld-profiles}
	
\begin{figure}
		\includegraphics[width=.5\textwidth]{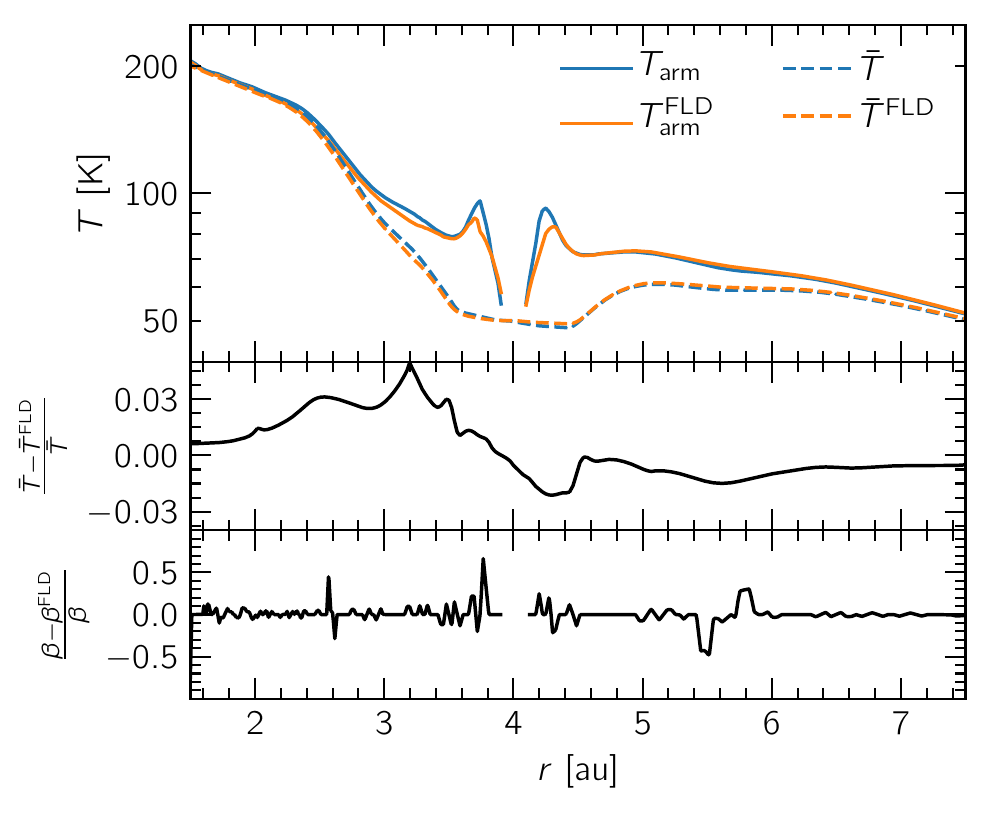}
		\caption{Comparison between two variants of our fiducial model ($M_\mathrm{p}=100$\,$\mathrm{M}_\oplus$), with and without radiative diffusion enabled. The effect of diffusion is visible near steep gradients along spiral crests or the gap edges, but negligible in general. Spiral arms in both models overlap exactly, as shown in the bottom panel.}
		\label{fig:fld-profiles}
	\end{figure}

	A comparison is plotted in Fig.~\ref{fig:fld-profiles} for the high-mass case. Spiral arms show slightly lower temperature maxima and the azimuthally averaged temperature profile is overall smoother, with lower highs and higher lows. This effect is strongest around parts of the disk that might contain steep temperature gradients, such as the region between 2.5--3.5\,au for these models, but still barely makes a difference of more than 3\% with respect to the model where we did not account for radiative diffusion, leading to identical aspect ratios in the two models and therefore practically indistinguishable pitch angles along spirals. We note that, for the 10-Earth-mass case, differences between the two models are much smaller. 
	
	\subsection*{Location and shape of the iceline}
	\label{appendix:fld-iceline}
	
	Temperature gradients are slightly different when accounting for radiative diffusion, and especially so around the gap edge. Since the iceline is relatively close to said gap edge in the two models where the module is enabled, we are more or less looking at the effects of radiative diffusion at its maximum potential. However, in the previous paragraph we found that its effect barely changes the picture with regard to shock strength, gap width or spiral location. Because of these three points, the iceline's location over time is expected to be slightly but not significantly different when compared to that found in our standard simulations.		
		
\begin{figure}[h]
		\centering
		\includegraphics[width=.5\textwidth]{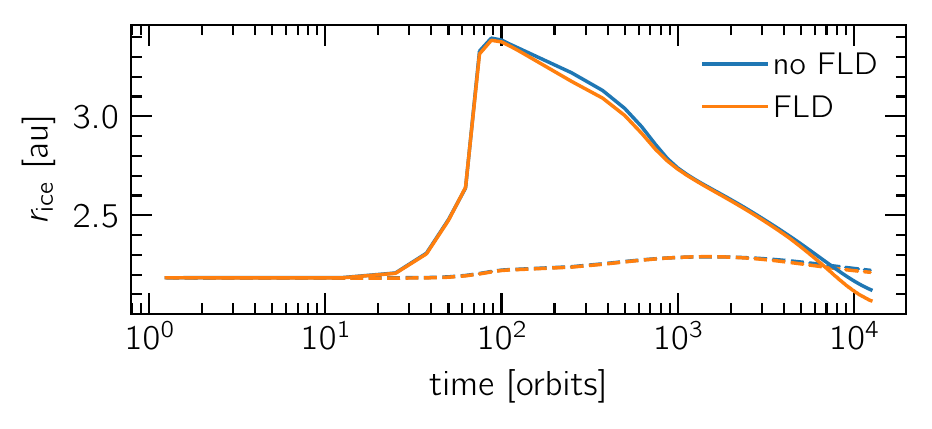}
		\caption{Evolution of the iceline's location for the two pairs of simulations with and without treatment of radiative diffusion. The solid and dashed lines refer to models where $M_\mathrm{p}=10$ and $100$\,$\mathrm{M_\oplus}$, respectively.}
		\label{fig:raddiff-icelines}
	\end{figure}

	In Fig.~\ref{fig:raddiff-icelines} we compare the location of $\bar{r}_\mathrm{ice}$ over time between our standard models and their respective FLD-enabled models. Indeed, the softer temperature profile in the inner disk allows the iceline to recede slightly more inwards when radiative diffusion is enabled. Nevertheless, the effect is still minuscule for the high-mass case and negligible for the low-mass case.
	
	\begin{figure}[h]
		\centering
		\includegraphics[width=.5\textwidth]{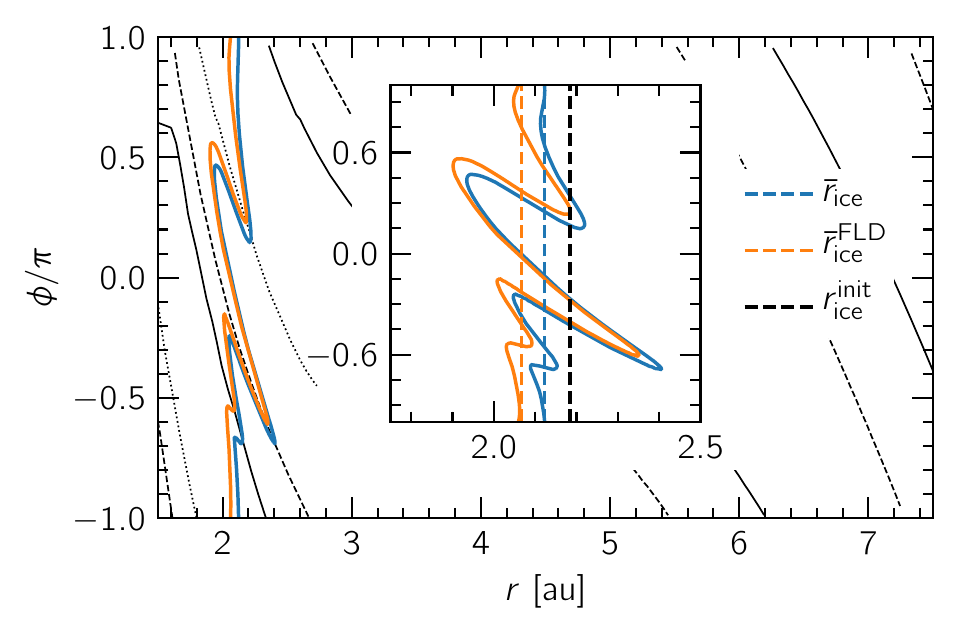}
		\caption{Shape of the water iceline at equilibrium for two models with and without radiative diffusion ($M_\mathrm{p}=100$\,$\mathrm{M}_\oplus$). Aside from the slight inward shift of the iceline, its shape is overall unaffected. We plot an extended range of our simulation for context. Solid, dashed and dotted black curves mark the location of primary, secondary and tertiary spirals, respectively.}
		\label{fig:raddiff-iceline-shape}
	\end{figure}

	A 2D analysis of the iceline's \emph{shape} returns similar results. As shown in Fig.~\ref{fig:raddiff-iceline-shape}, accounting for radiative diffusion does not change the shape of the iceline with respect to the standard model, but instead shifts it slightly inwards as shown in Fig.~\ref{fig:raddiff-icelines}. As with the 1D analysis, this difference is practically nonexistent for the low-mass case, as the temperature profile is overall softer: shocks by the 10-Earth-mass planet are significantly weaker and a gap does not open.

\section{Implementation of radiative diffusion}
\label{appendix:raddiff}

To examine the effect of radiative diffusion along the disk midplane, we implement an external module that couples to \texttt{PLUTO} and solves the following equation after every timestep:
\begin{equation}
\label{eq:fld}
\frac{\partial c_\mathrm{v}\Sigma T}{\partial t} = - \nabla\cdot\left(2H\bm{F}_\mathrm{rad}\right),
\qquad
\bm{F}_\mathrm{rad}=-\frac{4\sigma_\mathrm{SB}}{\lambda\rho\kappa} \nabla T^4,
\end{equation}
where $\bm{F}$ denotes the radiation flux across the disk midplane and is defined as:
\begin{equation}
\label{eq:rad-flux}
\bm{F}_\mathrm{rad}=-\frac{4\sigma_\mathrm{SB}}{\lambda\kappa\rho_\mathrm{mid}} \nabla T^4,
\end{equation}
where $\lambda$ is a flux limiter, following \citet{kley-1989}.

By defining a diffusion coefficient $K$ as
	\begin{equation}
	K=2H\frac{4\sigma_\mathrm{SB}}{\lambda\rho\kappa}(4T^3) = \frac{32\sigma_\mathrm{SB}\sqrt{2\pi}}{\lambda\Sigma\kappa}H^2T^3,
	\end{equation}
	we discretize Eq.~\eqref{eq:fld} following Appdx.~A.1 in \citet{muller-2014}
	\begin{equation}
	\label{rdiff-solver-breakdown}
	c_\mathrm{v}\Sigma\frac{\partial T}{\partial t}  =\frac{1}{r}\DP{}{r}\left(rK\DP{T}{r}\right)+\frac{1}{r^2}\DP{}{\phi}\left(K\DP{T}{\phi}\right),
	\end{equation}
	on a grid where $i$ and $j$ denote cell indices along the $r$- and $\phi$-direction respectively, and obtain:
	\begin{equation}
	\begin{split}
	&c_\mathrm{v} \Sigma_{ij} \frac{T_{i,j}^{n+1}-T_{i,j}^n}{\Delta t}=\\
	&\phantom{+}\frac{1}{r_i \Delta r_i}\left((rK)_{i+\frac{1}{2},j}\frac{T_{i+1,j}^{n+1}-T_{i,j}^{n+1}}{r_{i+1}-r_i} - (rK)_{i-\frac{1}{2},j}\frac{T_{i,j}^{n+1}-T_{i-1,j}^{n+1}}{r_{i}-r_{i-1}}\right)\\
	&+\frac{1}{r_i^2 \Delta \phi^2}\left(K_{i,j+\frac{1}{2}}(T_{i,j+1}^{n+1}-T_{i,j}^{n+1}) - K_{i,j-\frac{1}{2}}(T_{i,j}^{n+1}-T_{i,j-1}^{n+1})\right),
	\end{split}
	\end{equation}
	where $n$ and $n+1$ denote the states at time $t$ and $t+\Delta t$, respectively.
	
	We now have to solve for $T^{n+1}$. We can group up the right hand side to form a linear system:
	\begin{equation}
	\begin{split}
	T_{i,j}^n&=A_{ij}T_{i-1,j}^{n+1}+C_{ij}T_{i+1,j}^{n+1}+D_{ij}T_{i,j-1}^{n+1}+E_{ij}T_{i,j+1}^{n+1}+B_{ij}T_{i,j}^{n+1}\\
	&\Rightarrow \tensor{M}\cdot T^{n+1}=T^n,
	\end{split}
	\end{equation}
	where
	\begin{equation}
	\begin{split}
	A_{ij}&= -\frac{\Delta t}{c_\mathrm{v} \Sigma_{ij}}\frac{1}{r_i \Delta r_i}\frac{(rK)_{i-\frac{1}{2},j}}{r_{i}-r_{i-1}},\qquad
	C_{ij}= -\frac{\Delta t}{c_\mathrm{v} \Sigma_{ij}}\frac{1}{r_i \Delta r_i}\frac{(rK)_{i+\frac{1}{2},j}}{r_{i+1}-r_i},\\
	D_{ij}&=-\frac{\Delta t}{c_\mathrm{v} \Sigma_{ij}}\frac{1}{r_i^2 \Delta \phi^2}K_{i,j-\frac{1}{2}},\qquad
	E_{ij}=-\frac{\Delta t}{c_\mathrm{v} \Sigma_{ij}}\frac{1}{r_i^2 \Delta \phi^2}K_{i,j+\frac{1}{2}},\\
	B_{ij}&=1-A_{ij}-C_{ij}-D_{ij}-E_{ij},
	\end{split}
	\end{equation}
	are elements of the matrix $\tensor{M}$. We solve this system using successive overrelaxation (SOR). Therefore, we calculate and fix $\tensor{M}$ before iterating over:
	\begin{equation}
	\begin{split}
	\tilde{T}_{i,j}^{k+1}&=(1-\omega)\tilde{T}_{i,j}^{k}\\&-\frac{\omega}{B_{ij}}\left[A_{ij}\tilde{T}_{i-1,j}^{k}+C_{ij}\tilde{T}_{i+1,j}^{k}+D_{ij}\tilde{T}_{i,j-1}^{k}+E_{ij}\tilde{T}_{i,j+1}^{k}-T_{i,j}^{k=0}\right],
	\end{split}
	\end{equation}
	with $\omega=1.5$ until $\tilde{T}_{i,j}$ converges. Boundary conditions during this iterative process are closed, in order to conserve total thermal energy through the radiative diffusion substep.

\end{appendix}
\end{document}